# Response to Wright et al. 2018: Even More Serious Problems with NEOWISE

Nathan P. Myhrvold[1]

*Intellectual Ventures, Bellevue, WA 98005 USA*

## Abstract

Wright et al. (2018, hereafter W2018) responded to some of the irregularities and criticisms of the NEOWISE results raised by Myhrvold (2018a, 2018b) in a preprint that Wright et al. submitted for publication to *Icarus*. A response to the Wright preprint was also posted and submitted to *Icarus,* where the two manuscripts underwent joint peer review by the same reviewers. After many months and two rounds of reviewer comments, W2018 was withdrawn from consideration for publication. The present work is a version of the response manuscript to W2018. It has been amended to respond to all comments received from both rounds of peer review at *Icarus*. This response has also been expanded to address points raised by Wright et al. during the review process and in separate publications.  Though the withdrawal of W2018 precludes publication at *Icarus*, the response here presents important issues of broad interest to the planetary science community.

In their preprint, Wright et al. acknowledge the validity of some of the most serious issues identified by Myhrvold. Among these is the importance of Kirchhoff's laws and the copying of diameters from previously published analyses of radar, occultation, and spacecraft observations (collectively, ROS diameters), which were presented as the results of NEOWISE thermal modeling. While W2018 confirms that copying of ROS diameters occurred for 117 asteroids, it does not provide the actual modeled diameters or identify the asteroids for which copied diameters were substituted. W2018 confirms that many NEOWISE results represent curves that do not pass near the data points they claimed to fit.

In addition, W2018 reveals even more serious problems with the NEOWISE results than had previously been understood. Wright et al. explain for the first time that errors in the results were due to a software bug that was found and corrected in 2011 but that the team did not disclose until now. This bug apparently corrupted the vast majority of NEOWISE results, which have yet to be corrected. W2018 argues that the bug had only rare and small impacts on results, however W2018 elsewhere provides analysis that directly contradicts this claim.

Other arguments offered in W2018 to rebut claims of Myhrvold (2018b, hereafter M2018b). We learn from W2018 that $H$ values used by NEOWISE were not taken from the literature, as previously claimed, but were in fact derived by fitting data, a method that is inconsistent with prior NEOWISE papers and previous practice in thermal modeling. The systematic underestimation of *WISE* observation errors shown in M2018b is confirmed here with a new analysis of repeated observation of calibration stars.

---

[1] nathan@nathanmyhrvold.com

For each argument in the response of Wright et al., I show that W2018 is mistaken. Ultimately, W2018 fails to make valid arguments supporting any of its criticisms of M2018b, and instead reveals why the results of the empirical examination that Myhrvold published were important.

# 1. Introduction

In Wright et al. (2018, hereafter W2018), the NEOWISE team responds to some of the criticisms made by two recent papers that examined different aspects of asteroid thermal modeling (Myhrvold 2018a, 2018b). Although W2018 was withdrawn from its submission to *Icarus* after both it and my reply here had been peer reviewed, the W2018 preprint remains on arXiv, and it raises additional issues and present additional evidence that warrant examination and discussion.

Myhrvold (2018a) presented an analysis of thermal modeling in the case when observations are made in bands containing a substantial amount of reflected sunlight. In such cases, Kirchhoff's law of thermal radiation governs the relation of emission to reflection. Many prior thermal-modeling studies employed observations in long-wavelength infrared that are negligibly affected by reflected sunlight and could thus ignore Kirchhoff's law of thermal radiation. In contrast, the *WISE*/NEOWISE mission (Wright et al., 2010) uses observations from the *WISE* spacecraft in two bands (W1 and W2) that include substantial reflected sunlight. The NEOWISE modeling approach explicitly violated Kirchhoff's law (Myhrvold 2018a).

The primary focus of Myhrvold (2018a) was to examine the importance of Kirchhoff's law for asteroid thermal modeling and to derive a version of the NEATM that would satisfy the law. In addition, the paper examined the modeling approaches performed by NEOWISE and reported results from Monte Carlo simulations that assessed the potential impact of the violation of the law on the accuracy of physical parameters estimated from the model in certain cases. W2018 describes its Equation (4) as "designed to satisfy Kirchhoff's law," thus acknowledging the importance of Kirchhoff's law in thermal modeling, and it also does not dispute that NEOWISE models violate Kirchhoff's law.

Myhrvold (2018b, hereafter M2018b) empirically analyzed the NEOWISE results across multiple papers (Grav et al., 2011a, 2011b, 2012; Mainzer et al., 2011c, 2011a, 2011b, 2014, 2016; Masiero et al., 2011, 2012, 2014). Together, these results include estimates of physical characteristics for more than 130,000 asteroids. The empirical analysis in M2018b, which focused primarily on results from the fully cryogenic portion of the mission, identified numerous inconsistencies and technical problems across the NEOWISE papers that raise serious concerns about the reliability of those results.

The response of the NEOWISE team to the issues I have identified is welcome because the NEOWISE results hold enormous potential scientific value. This data set, vastly larger in scope than those produced by prior studies, represents the largest collection of estimated physical parameters of asteroids (particularly diameters) yet produced. These parameters contribute to our understanding of the solar system and its formation, as well as to practical issues such as planning planetary defense against asteroid impacts.

The NEOWISE observational data set also offers invaluable opportunities to apply and compare different modeling techniques—but only if researchers throughout the community fully understand how the data were collected and processed. To date, many critical details necessary to fully utilize the data have not been provided. The response by Wright et al. now sheds light on some of those details. Further elucidation of others is needed.

W2018 now confirms several of the most serious flaws with NEOWISE that were identified by M2018b. Among these is the fact that diameter estimates for 117 asteroids presented in result tables labeled as NEOWISE thermal-modeling results were instead copied without disclosure from prior radar, occultation, and spacecraft (ROS) studies. W2018 does not present the modeled results for these 117 asteroids, nor does it identify the objects.

M2018b found that about 49% of NEOWISE model fits did not actually fit the data they were purported to model, as the fit curves missed entire bands of data. W2018 now confirms that these errors arose from a bug in NEOWISE software that was found and fixed in 2011. The bug corrupted thousands of diameter estimates published in Masiero et al. 2011, and possibly in other NEOWISE papers published in 2011. W2018 represents the first published notice of the bug.

M2018b showed that the 2016 PDS archive contained numerous irregularities among NEOWISE entries. These included the deletion or addition of thousands of asteroids, the modification of thousands of results, and false attribution of modified results to earlier NEOWISE papers. In particular, all of the asteroid results having copied ROS diameters in Masiero et al. 2011 were deleted from the PDS submission.

Although the 2016 PDS submission modified thousands of results, those changes did *not* include updates to diameter results that were corrupted by the software bug. The NEOWISE group was independently informed of this issue in 2015 by a draft version of the M2018b analysis. The bug-corrupted diameter estimates were nevertheless submitted to the PDS in 2016. No subsequent NEOWISE papers included corrected data or errata.

W2018 now acknowledges those irregularities but minimizes their importance as "some minor issues with consistency between tables due to clerical errors in the *WISE*/NEOWISE team's various papers and data release in the Planetary Data System."

W2018 also discloses for the first time a linear correction to the W3 band that was mentioned in NEOWISE papers but was incorrectly referenced to the *WISE* Explanatory Supplement (Cutri et al., 2011). The origin or justification of the formula is not presented in W2018, and as discussed below in section 2.3, it raises further questions about the foundational NEOWISE studies.

W2018 confirms the finding of M2018b that the NEOWISE results violate the relationship between absolute magnitude $H$ and $D$ and visible-band albedo $p_v$. W2018 shows that this occurs because the NEOWISE papers treated $H$ as a free parameter to be fit, rather than as an input to the analysis. This approach was not disclosed in any prior NEOWISE paper.

The resolution over debate about the occurrence of the issues above—if not about their severity and remedy—represents real progress in the ongoing controversy about NEOWISE because, in these cases at least, they end the debate over *whether* these things happened. We can instead focus on the exculpatory rationale presented in W2018 and whether it adequately explains why they happened.

In another set of issues raised by M2018b, however, Wright et al. in W2018 stridently object to my analysis and offer technical arguments to support their position. I show in this reply that none of the counterarguments offered in W2018 refute or effectively rebut the findings of M2018b. To the contrary, W2018 highlights the salience of the critique of M2018b and the need to resolve many questions regarding NEOWISE that remain unanswered.

## 2. Calibration issues

*2.1 NEOWISE "Calibration" and Error Analysis*

In 2011, the NEOWISE group published two papers in the *Astrophysical Journal* (ApJ) that are central to several of the issues covered in M2018b and W2018. The first paper was "Thermal Model Calibration for Minor Planets Observed with *Wide-Field Infrared Survey Explorer*/NEOWISE," (Mainzer et al., 2011b, hereafter ApJ 736). The second was "Thermal Model Calibration for Minor Planets observed with *WISE*/NEOWISE: Comparison with *Infrared Astronomical Satellite*" (Mainzer et al., 2011c, hereafter ApJ 737).

These two papers, and ApJ 736 in particular, were cited by subsequent NEOWISE papers to justify repeated claims of "±10%" accuracy, with varying and often contradictory qualifiers (detailed in section 7). As detailed in M2018b, ApJ 736 mentions 117 objects (mostly asteroids, but also a few moons) for which diameter estimates had been obtained from ROS observations. That paper studied 50 of those objects (48 asteroids and two moons) in detail. Referring to ApJ 736 as Mainzer et al. (2011a), W2018 states that:

> Asteroids have red spectral energy distributions (SEDs), and the *WISE* band passes are broad. Mainzer et al. (2011a) sought to verify that the zero points and color corrections derived for the *WISE* band passes from calibrator stars (which are blue) and Active Galactic Nuclei (which tend to be red but can be variable) were appropriate for objects with very red SEDs such as asteroids. To that end, the differences between model and observed magnitudes were plotted vs. the asteroid calibrator objects' sub-solar temperatures when their diameters were held fixed to previously published values in order to verify that the newly derived color corrections worked properly for these asteroids, which are much cooler than stars.

Although the title of ApJ 736 claimed "thermal model calibration," the NEATM has no overall calibration parameters that can be set. Instead, the NEATM makes use of the beaming parameter $\eta$, which is fit independently for each application of the model (Harris, 1998). The only kind of calibration possible when using the NEATM is photometric—i.e., setting the zero points for *WISE* magnitude. ApJ 736 did not describe photometric calibration as part of its methodology.

Photometric calibration for *WISE* was performed by Jarett et al. (2011), in a study that used thousands of calibration objects located at the ecliptic poles, where *WISE* observes each source once per orbit and thus obtains sufficient statistical power to perform a reasonable calibration and to check for drift.

In contrast, ApJ 736 compared a computed model flux for 50 objects to the observed flux, though typically for only *one observation per object* (Table 2 in ApJ 736). The proper procedure to calibrate zero points and color corrections is to *vary* those zero points and/or color corrections and to then determine whether values from Wright et al. (2010) are optimal. Instead, the authors of ApJ 736 concluded that "Nevertheless, most of the predicted magnitudes are in good agreement with the observed magnitudes, indicating that the procedure given in Wright et al. (2010) for color correction is adequate." As M2018b noted, this describes a very rough consistency check; it is not appropriate to rely on it as a true calibration. W2018 does not challenge this point.

Although ApJ 736 failed to perform a true calibration, subsequent NEOWISE studies relied on it being an accuracy analysis that can be applied to all NEOWISE thermal modeling. Though that was apparently one goal of the ApJ 736 paper—its Table 1 listing results for the 50 objects is titled "Spherical NEATM Models were Created for 50 Objects Ranging from NEOs to Irregular Satellites in Order to Characterize the Accuracy of Diameter and Albedo Errors Derived from NEOWISE Data"—ApJ 736 fixed the diameter to a ROS value and then calculated fluxes, a procedure that is the reverse of the modeling approach described in the results papers. The procedure used in ApJ 736 must have involved some fitting, because the beaming parameter $\eta$ is determined by fitting. Fitting could also have been used to compute modeled diameters.

The procedure used in ApJ 736 is thus not suitable for use either as characterization of accuracy or as an error analysis (M2018b). Moreover, ApJ 736 analyzed asteroids having ROS diameters, which are typically large objects that tend to saturate the sensor, particularly in the W3 band. Conclusions about the zero-point "calibration" drawn from analysis of saturated observations are of questionable validity for the vast majority of NEOWISE asteroids, which are not saturated.

After estimating the flux differences (Table 2 of ApJ 736), the authors claimed that:

> Since diameter is proportional to the square root of the thermal flux (Equation (1)), the minimum systematic diameter error due to uncertainties in the color correction is proportional to one-half the error in flux. These magnitude errors result in a minimum systematic error of ~5%–10% for diameters derived from *WISE* data; they are of similar magnitude to the diameter uncertainties of most of the underlying radar and spacecraft measurements, which are ~10% (references are given in Table 1).

M2018b explains at length why this assumption is invalid. The flux *in a single band* is indeed proportional to the square of the diameter, *assuming everything else is held constant*. However, NEATM modeling employs *multiple* bands, and unless the magnitude in each band varies by *exactly the same amount*, changes in total flux do not have that simple relationship to diameter. In fact, the relative amounts of flux in each band help determine the shape of the NEATM SED, and thus the sub-solar temperature $T_{ss}$. One thus must perform the NEATM nonlinear least-squares model fit to assess the impact (Myhrvold 2018a, 2018b).

Note that because the observed and modeled fluxes given in Table 2 of 736 have different means and standard deviations, there are strong empirical reasons to believe that the fluxes do not vary in exactly the same way in all bands. If that is the case, then diameter estimates for these 50 objects could *not* simply be half of the corresponding flux uncertainty. Absent thermal modeling, it is not possible to determine whether the zero-point fluxes in Wright et al. 2010 were consistent or not.

W2018 does not refute the detailed explanation of this issue in M2018b, yet it repeats the same error of ApJ 736 by claiming incorrectly that "Since an asteroid's flux goes as the square of its diameter, the minimum diameter uncertainty must scale as one-half of the flux uncertainty."

Because the ApJ 736 paper provides neither a valid photometric calibration nor a valid error and accuracy analysis for NEOWISE, we must question the numerous claims of ±10% accuracy that reference it as such (see section 7).

*2.2 Copying of ROS diameters*

The flaws and interpretation of ApJ 736 results are relevant to the present debate because of how these asteroids were treated in the principal NEOWISE result papers, notably by Masiero et al. (2011), which presented results for main-belt asteroids, and by Mainzer et al. (2011c), which presented results for near-Earth asteroids. M2018b checked the tables of NEOWISE thermal-modeling results in those two papers against previously published ROS diameters and observed that in cases involving more than 100 asteroids, estimates listed in these NEOWISE papers as modeled diameters matched estimates from previously published ROS studies exactly to within the precision quoted (1 m).

Because the only way to assess the accuracy of the NEOWISE model estimates is to compare them to independent estimates—i.e., to ROS diameters—the copying of ROS diameters into these tables in Masiero et al. (2011) and Mainzer et al. (2011c) prevented others from assessing the accuracy that NEOWISE modeling provided for *any* of the asteroids with results in those studies.

ApJ 737 compared ROS diameters to diameter estimates obtained by Ryan and Woodward (2010) and by thermal modeling from *IRAS* (Tedesco et al., 2002). On the basis of that comparison, ApJ 737 was sharply critical of those earlier studies.

Regarding the absence of a comparison to ROS diameters in ApJ 736, W2018 argues that:

> Since the point of the analysis in Mainzer et al. (2011a) was to verify the then-newly derived zero points and color corrections for the four *WISE* bands, a plot of the previously measured diameters of the calibration objects vs. the diameters derived for the objects using *WISE* data was not shown for that set of observations.

W2018 does not explain why the ApJ 737 paper failed to do so, nor why Wright et al. do not present the model-derived diameters in W2018.

Regarding the inclusion of copied ROS diameters in the results papers, W2018 states:

> We note that in the process of compiling the results for Table 1 of Masiero et al. (2011) and Table 1 of Mainzer et al. (2011b), the full list of calibrator objects was incorporated in an effort to be consistent with Mainzer et al. (2011a). However, late in the referee process for Mainzer et al. (2011a), the referee requested that objects with maximum lightcurve amplitudes larger than 0.3 mag be dropped from that paper, as these are more likely to be highly elongated and thus poor choices for calibrators. Neither Table 1 in Mainzer et al. (2011b) nor Masiero et al. (2011) was updated to reflect this reduced calibration set due to an oversight when the calibrator object table was reproduced in these two papers, affecting 0.7% of objects (3 objects) in Mainzer et al. (2011b) and <0.1% of objects (68 objects) in Masiero et al. (2011)

(Mainzer et al. 2011b in the passage above is the near-Earth asteroid results paper referenced here as Mainzer et al. 2011c.)

Wright et al. claim that copied ROS diameters were published instead of modeled diameters "to be consistent with" Mainzer et al. 2011a (i.e., ApJ 736). However, the goal of consistency would be best served by applying the NEOWISE model to estimate diameters for these asteroids in the same manner as done *for all other asteroids*. One could then make the same kind of comparison to ROS diameters that the ApJ 737 paper did for estimates from *IRAS* and Ryan and Woodward. Treating 117 asteroids differently from all the others is fundamentally *incompatible* with consistency. The inconsistent treatment was not disclosed by either Masiero et al. (2011) or Mainzer et al. (2011c).

Repeated requests by me and other researchers for the NEOWISE group to disclose the identity and NEOWISE-modeled diameters of these 117 asteroids remain unfulfilled. M2018b inferred and reported the names of 105 of the objects, but these are not confirmed by W2018. At least 12 cases remain unidentified. Figure 6 of W2018 compares *IRAS* diameter estimates with NEOWISE estimates, both before and after correction of a bug in the NEOWISE code (section 4.1). No comparison of this kind to ROS estimates is made for the 117 asteroids of special interest.

Of the 105 asteroids with copied diameters identified in M2018b, 87 are among the subset of 1700 asteroids that W2018 analyzes with both 2011 code and with code run in 2018 to produce its Figure 6(f). W2018 does not, however, list the diameters of those 87 ROS objects.

W2018 recounts a reviewer request as the explanation of why the ApJ 736 paper discusses 117 objects in its text yet lists only 50 objects in its Table 1. However, W2018 only cites the aforementioned "consistency" with the Ap 736 paper to explain why the copied diameters were included in the thermal modeling result tables of Masiero et al. (2011) and Mainzer et al. (2011c) W2018 does not state why those papers did not disclose and justify the inclusion of these copied diameters.

An earlier public explanation of those decisions was published by Joseph Masiero on the Yahoo minor planets forum in response to a question from an academic colleague (Jean-Luc Margot of UCLA). In that post (Masiero, 2016), Masiero stated that the team copied the ROS diameters because doing so allowed more accurate determination of $\eta$. He claimed that the practice was properly referenced:

> Regarding the diameters, as explained in the first thermal model calibration paper (Mainzer et al. 2011 ApJ 736, 100), the radar diameters were held fixed and were used to derive fluxes to verify the color correction calibration. We quote those fits in that paper, as well as the later MBA and NEO papers, both of which reference the 2011 ApJ 736, 100 paper in this respect. As described in this paper, if the radar or occultation diameters were available, they were used since they allow for improved solutions for the other free parameters such as beaming, and the thermal model calibration paper was cited (see e.g., Section 3 of Mainzer et al. ApJ 2011 743, 156). For more information, you can consult the 2011 ApJ 736, 100 paper.

These claims do not withstand close inspection and are entirely different than with W2018's claim of seeking "consistency" with the ApJ 736 paper. Instead, Masiero argues that the coping was done to improve estimates of $\eta$, and that the copying was properly disclosed and referenced. References given in the paper to ApJ 736 *do not* indicate that diameters were copied — i.e., referenced "in that respect" in the quote above. (See the Appendix for relevant excerpts from Masiero et al. 2011 and Mainzer et al. 2011c.)

ApJ 736, ApJ 737, and all subsequent NEOWISE papers have focused primarily on the accuracy of *diameter* estimates and secondarily on visible-band geometric albedo or near-infrared albedo. The abstract of Masiero et al. (2011), for example, claimed that "Using a NEATM thermal model fitting routine, we compute diameters for over 100,000 Main Belt asteroids from their IR thermal flux, with errors better than 10%." Nowhere in these papers does the group claim high accuracy for their solutions for the beaming parameter ($\eta$). It clearly would be of little use to improve $\eta$ for an *undisclosed* set of 117 asteroids hidden among actual thermal-modeling results for more than 100,000 asteroids.

Moreover, using copied diameters would not necessarily improve estimates of $\eta$. ApJ 736 itself argues that the error in NEOWISE diameter estimates is comparable to, or better than, the estimates for ROS-derived diameters—at least that is the implication, because both ApJ 736 and Masiero et al. 2011 claim that ROS diameters are not more accurate than their thermal modeling diameters.

M2018b raised the issue that the diameters appear to be copied, as well as concerns about improper and missing citation of prior works the NEOWISE papers. W2018 has confirmed that the diameters were copied. The attempted explanation of "consistency" is not a scientific argument, however, so it must be backed up with a more fulsome explanation. Masiero, a co-author of W2018, should also explain why his explanation has changed since 2016 and address the issue of disclosure and referencing.

## 2.3 Saturation and the W3 correction formula

The brief section in W2018 titled "The linear correction for non-linearity" provides an equation for making a correction to the W3 band when it may be saturated (below a threshold magnitude). As the authors do not explain why this is provided, how it is useful, or how it is relevant to M2018b, some context is provided here.

The earliest mention of the formula is in the ApJ 736 paper, which states:

> Objects brighter than W3 = 4 and W4 = 3 mag were assumed to have flux errors equivalent to 0.2 mag due to changes to the shape of the point-spread function as the objects became saturated, and a linear correction was applied to the W3 magnitudes in this brightness regime (the *WISE* Explanatory Supplement contains a more detailed explanation).

An identical sentence is found in the ApJ 737 paper, and similar text is found in most NEOWISE papers. The passage implies that the linear correction is a general photometric adjustment for the *WISE* telescope and associated pipeline found in the *WISE* Explanatory Supplement (Cutri et al., 2011), which is not specific to asteroids or the NEOWISE project.

However, this is not the case. I was unable to find the correction in (Cutri et al., 2011) or to obtain it from the *WISE* help desk. The first author, Roc Cutri, responded to me that he was unaware of such a correction. In response to an email inquiry, *WISE* principle investigator, NEOWISE coauthor, and W2018 first author Edward Wright wrote that he knew nothing about the linear correction, and he agreed that he could not find it in the Explanatory Supplement. He referred me to a paper on AGB stars that recommended a slight correction in bands W1 and W2 but not W3 (Lian et al., 2014).

This correction formula—one of a number of missing details that has so far prevented replication of the NEOWISE results—is important because ROS asteroids tend to be heavily saturated, particularly in bands W3, W4. That is true for both the subset of asteroids used in the ApJ 736 paper and the larger set for which diameters were copied. Table 1 lists the counts and frequencies of asteroids for which one or more bands is saturated, meaning in this case that the observed magnitudes are below the following thresholds: $W1 < 6$, $W2 < 6$, $W3 < 4$, $W4 < 3$, a definition consistent with that used in ApJ 736 and the *WISE* Explanatory Supplement.

|  | **Asteroid Counts and Percentage** | | | | | | |
|---|---|---|---|---|---|---|---|
| **Data Set** | Total | W1 | W2 | W3 | W4 | At least 1 band | No saturation |
| ApJ 736 | 46 | 0 | 1 | 37 | 38 | 38 | 8 |
|  | 100% | 0% | 2.17% | 80.4% | 82.6% | 82.6% | 17.4% |
| Copied ROS asteroids | 105 | 0 | 5 | 85 | 93 | 93 | 12 |
|  | 100% | 0% | 4.76% | 81.0% | 88.6% | 88.6% | 11.4% |
| All asteroids in fully cryogenic phase | 147,681 | 3 | 9 | 542 | 1321 | 1324 | 147,357 |
|  | 100% | 0.002% | 0.006% | 0.37% | 0.089% | 0.09% | 99.1% |

**Table 1.** Counts and percentage distribution across different data sets of asteroids with one or more bands saturated. The ApJ 736 set were analyzed in Mainzer et al. 2011a. The copied ROS asteroids are the subset of 105 asteroids identified in M2018b of the 117 asteroids for which NEOWISE papers reported a copied ROS diameter estimate as a NEOWISE thermal-modeling result. The "all asteroids" data set includes all asteroids observed during the fully cryogenic phase of the *WISE* mission.

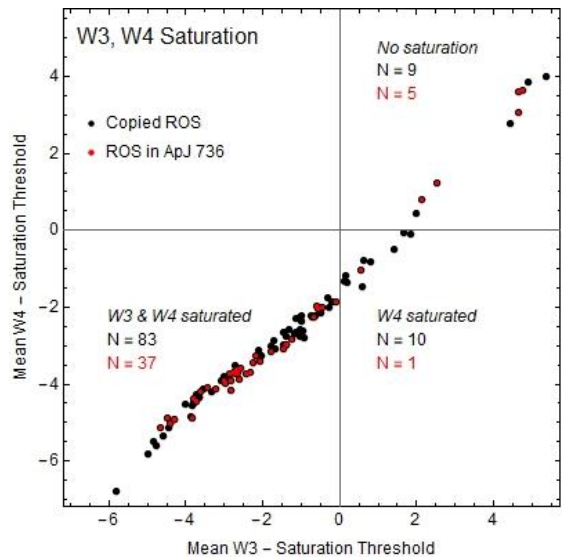

**Figure 1. Mean magnitudes in W3, W4 minus the saturation threshold plotted for ROS asteroids.** Asteroids plotted are those identified in the ApJ 736 paper (red points) and 105 asteroids for which diameters were copied from ROS sources (M2018b, black points). Negative values indicate the distance from the mean magnitude in the band to the saturation threshold (in magnitudes). Labels in each quadrant show the saturation state and counts for the data sets of asteroids with ROS diameter used by NEOWISE.

Almost all (88.6%) of the asteroids for which ROS diameter estimates were available are saturated in at least one band, and usually in both the W3 and W4 bands (Table 1). In contrast, only 0.09% of the overall population of asteroids observed by *WISE* are saturated in one or more bands.

The degree to which the observed fluxes are brighter than the saturation is shown in Figure 1, which plots the mean magnitude in W3 and W4 minus the saturation threshold. Negative values indicate saturation. Cases with saturation in both the W3 and W4 bands extend out more than 6 magnitudes brighter than the threshold. Examination of the observational data shows that saturation occurs for nearly every data point in the affected bands, with only a few exceptions.

The physical explanation for the extensive degree of saturation found in asteroids with ROS diameters is that they tend to be large, close to Earth, or both. These factors make them accessible to ROS observation but also typically make them bright enough to saturate the *WISE* sensors.

Because saturation is so widespread in this data set, the correction formula for W3 presented in W2018 is of critical importance to understanding the role of "calibration" in the ApJ 736 paper. For the asteroids saturated in these bands, the "correction" essentially determines or heavily influences the resulting diameter estimate

A recent poster by Wright (Wright, 2019) derives a W3 saturation formula by adjusting its parameters to achieve zero bias in the diameter estimates for ROS asteroids. In effect, Wright uses thermal modeling as a means to calibrate a linear correction formula for W3 in the saturated regime. Wright reports that:

> This analysis uses a different approach than the original NEOWISE calibration paper (Mainzer et al. 2011, ApJ 737, L9) but obtains the same slope: 0.855 here vs 0.86 for Mainzer et al.

We learn from this poster that the linear correction formula was actually a *result* of analysis for the NEOWISE "calibration" paper(s) (i.e., ApJ 736 and ApJ 737) and not an input to these papers from other *WISE* calibration work. W2018 notes that the approach presented in the paper is "different" but does not disclose what approach was used in in the NEOWISE papers.

Taken together, W2018 and the poster imply that correction formula was part of the "calibration paper," which explains several other unusual aspects of these papers. As discussed above, the title referring to "thermal model calibration" makes little sense if the goal was to check the *WISE* zero points—it was a rough consistency check, not a calibration. But the term "calibration" *would* make sense if the goal of the papers was to use thermal modeling to derive the W3 saturation formula. Doing so would involve a true calibration that would determine slope and intercept constants—not merely the rough consistency check described in section 2.1. In such a procedure, the asteroids would serve as "calibrators," as W2018, ApJ 736, and ApJ 737 refer to them, because they would have been used to create or adjust the linear correction formula.

Neither ApJ 736 nor ApJ 737 disclose that they originated or derived the W3 saturation correction formula, however. Given the remarks in those papers that the observed flux was close to the modeled flux, one would expect a disclosure that observed fluxes were altered to approximate modeled fluxes by applying a correction formula.

There are many differences between the analysis in Wright 2019 and the analysis used in NEOWISE papers. These may or may not account for the "different approach" mentioned in the quoted passage. Wright 2019 uses a undisclosed set of 158 asteroids with ROS diameters; that set clearly differs from the 50 or 117 asteroids used in ApJ 736. The NEOWISE papers use the NEATM, but Wright 2019 uses a modified form of the NESTM (Wolters and Green, 2009), without explaining that choice. NESTM has a free parameter that is set to a value, which is not explained in the poster. Nor is there a sensitivity analysis of how that free parameter impacts the slope of the linear correction, and whether that seemingly arbitrary choice explains the linear correction slope of 0.855.

Neither Wright (2019) nor W2018 addresses the role of saturation in bands other than W3. Wright (2019) presents a calibration formula for W2 based on a comparison between *WISE* and 2MASS observations of stars, but W2 band is rarely saturated in *WISE* observations of asteroids (Table 1), but saturation of the W4 band is even more common than it is for W3, and very often W4 and W3 are saturated in the same observation.

A correction that adjusts the saturated W3 flux value without similarly adjusting a saturated W4 flux value will alter the relative contribution of each. Fitting the NEATM SED in such cases will thus yield a different sub-solar temperature $T_{ss}$. As demonstrated in Myhrvold (2018a), a small relative shift can make a large impact on estimated diameter.

The NEOWISE claims of ±10% accuracy in diameter originate from analysis of asteroids with ROS estimates studied in the ApJ 736 paper, which are a highly saturated sample. An accuracy measurement for those samples, *even if made correctly*, would not be generalizable to the vast majority of the NEOWISE asteroids, which are not saturated (Table 1). Wright (2019) found that applying an improved correction formula still leaves a residual 13.3% scatter at the 1σ level with respect to ROS diameters. So even after adjusting diameter bias, considerable random scatter remains.

## 3. Comparing the accuracy of multiple estimates

### 3.1    NEOWISE accuracy comparison in the literature

Contrary to the claims made in W2018, there is no single "correct" or optimal way to make statistical comparisons among multiple estimates of an asteroid's diameter, where each estimate involves an associated uncertainty. Several methods have been used to compare NEOWISE estimates to ROS diameter estimates and to estimates from other thermal modeling studies such as *IRAS* and AKARI.

Such comparisons are complicated by the fact that each estimate has an associated uncertainty; the comparison is therefore statistical. Moreover, when multiple NEOWISE or ROS estimates are available for an asteroid, this must also be taken into account. Each comparison method has features that may make it desirable or undesirable in a given context.

The context relevant here to the choice of method of comparison is our goal of assessing the accuracy of NEOWISE more generally. Although ROS diameter estimates are only a tiny subset of the overall population of asteroids (for NEOWISE, 117 out of >130,000 objects), they are the only independent points of comparison available. *IRAS* and AKARI estimates are available for a larger set of objects but that set is still much smaller than the overall NEOWISE population.

In general, the NEOWISE result papers from 2011 through the present do *not* perform numerical comparisons of NEOWISE estimates with other sources. Masiero et al. 2011 and subsequent NEOWISE papers claimed an accuracy of ±10% or better without providing supporting calculations and instead supplied only references to ApJ 736 (section 7). But the ApJ 736 paper did not compare diameter estimates at all (section 2.1). That paper did use ROS diameter estimates to compare observed and modeled fluxes. In cases where more than one ROS estimate was available, only one was used, and no process for selecting that estimate was provided. ROS estimates having the lowest error were not always selected. Estimates derived from particular publications or methods were not consistently preferred.

For example, ROS estimates were published for asteroids 208 and 64 by both Shevchenko and Tedesco, 2006 and Ďurech et al., 2010. ApJ 736 used the estimate from Ďurech et al. for asteroid 208 but the estimate from Shevchenko and Tedesco for asteroid 64. ApJ 736 contains many such inconsistencies (listed in M2018b Supplemental Information, Table S1).

The ApJ 737 paper plots diameters estimates from *IRAS* (Matson, 1986) and Ryan and Woodward (Ryan and Woodward, 2010) versus NEOWISE diameters or ROS diameters for asteroids those data sets have in common. But ApJ 737 contains no *numerical evaluation* of the accuracy. Thus the criticism in the paper of *IRAS* and RW estimates as more "biased" than NEOWISE is not a valid inference from evidence presented in ApJ 737.

Usui et al. (2014) published the first numerical comparisons of diameter estimates from AKARI, *IRAS*, and NEOWISE. For each asteroid, they calculated the ratio $D_x/\overline{D}$, where $D_x$ is the diameter estimate for the asteroid for $x$ = AKARI, *IRAS*, ROS and $\overline{D}$ is the mean of the diameter estimates for that object produced by the three studies. A normal distribution was then fit to the histogram of the $D_x/\overline{D}$ ratios of all of 1993 asteroids that were common to the three surveys.

W2018 criticizes M2018b on the basis that it "ignored" Usui et al. 2014. To the contrary, M2018b used AKARI data to compare diameter estimates from *IRAS*, ROS, Ryan and Woodward (Ryan and Woodward, 2010), and NEOWISE. M2018b did not adopt the method used in Usui et al. 2014 for several reasons. First, using each $D_x$ in both the numerator and denominator creates a correlation that tends to overstate the accuracy, a fact also noted by W2018.

Second, the method used by Usui et al. assumes that the ratios $D_x/\overline{D}$ have no error (section 3.2). Third, in the 75.2% of cases where multiple NEOWISE estimates were available for an asteroid, Usui et al. 2014 apparently averaged all of those estimates. This approach prevents the results from being generalizable to all asteroids in the NEOWISE data set, in which multiple diameter estimates are available for just 9.6% of asteroids (section 3.3). Finally, Usui et al. 2014 included asteroids from the 3-band portion of the NEOWISE mission along with asteroids from *IRAS* that were detected in only a single band.

Figure 2 and Table 2 show results that replicate the Usui method, and variations of it, and that illustrate some of the issues that M2018b avoided by using an improved method. Whereas Table 2 in Usui et al. 2014 presented their results as the parameters of a best-fit normal distribution, Table 2 and the top row of Figure 2 here report the 68.27% confidence intervals, which are more appropriate for the asymmetric distributions seen in these histograms.

Simple ratios of diameters are inadequate for uncertainty analysis in cases such as this where each of the diameter estimates has an associated uncertainty—a point discussed further in section 3.2. The row labeled "+ uncertainty" of Table 2 approximates the distribution with a Monte Carlo evaluation of the ratio $D_x/\bar{D}$ that uses 10,000 samples for both $D_x$ and $\bar{D}$. The row "+ don't average NEOWISE" includes these uncertainties and also includes all NEOWISE diameter estimates available for a particular asteroid rather than averaging them to produce a single mean estimate.

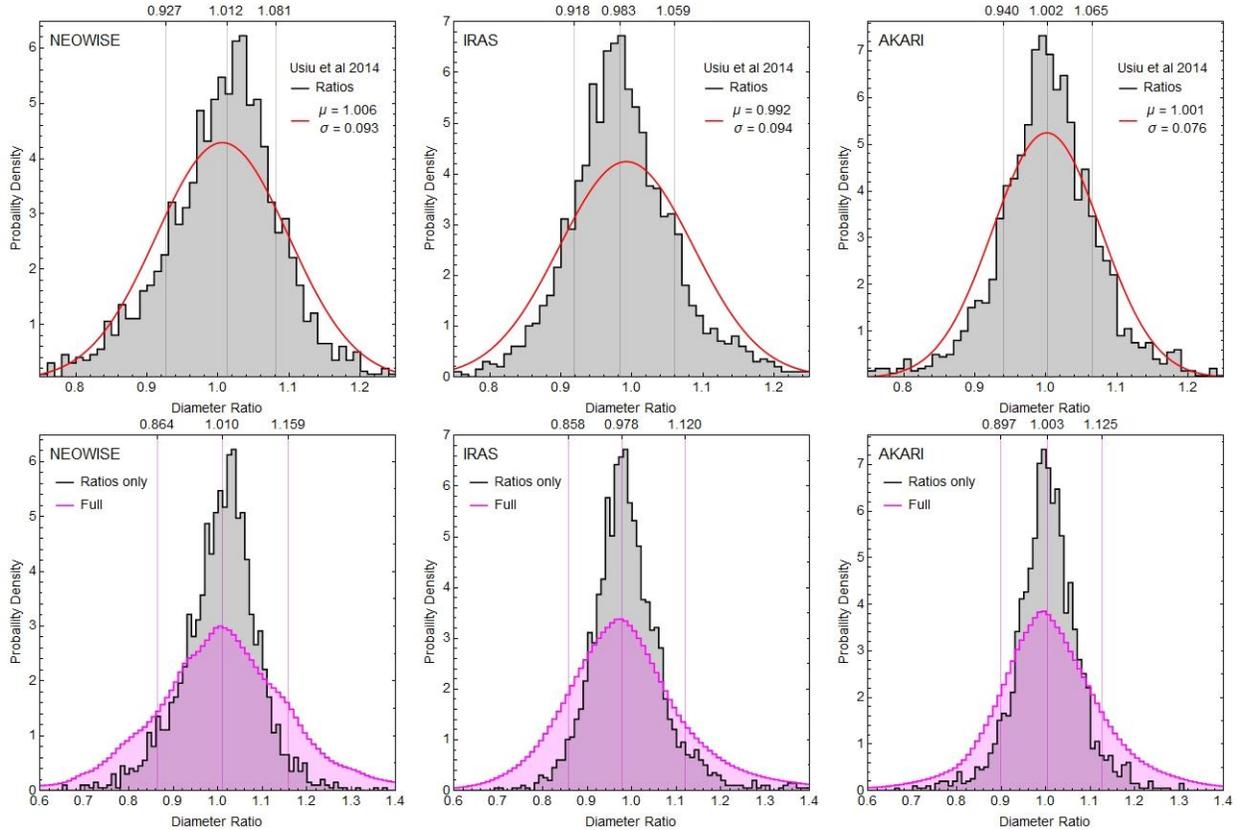

**Figure 2. Histograms from Usui et al. 2014 and variations.** Gray histograms were replicated by using the methodology of Usui et al. 2014 on diameter estimates for the 1993 asteroids that the *IRAS*, AKARI, and NEOWISE surveys have in common. Top row: histograms for the diameter estimates from the named source divided by the mean of the diameter estimates from all three surveys. Vertical lines identify the median and 68.27% confidence interval for each histogram. Red curves are best-fit normal distributions with the parameters published in Usui et al. 2014, Table 2. Due to asymmetry in distributions, not all normal distributions offer good fits to the histogram. Bottom row: histograms as above, overlaid by magenta histograms corrected to avoid autocorrelation and to account for uncertainty in the ratios. Vertical magenta lines mark the medians and 68.25% confidence intervals for the corrected distributions.

To avoid artificially exaggerating the accuracy by using a mean value $\bar{D}$ that correlates the numerator and denominator, we can redefine $\bar{D}$ as the mean of the two *other* sources, excluding the source being studied. To examine NEOWISE accuracy, for example, one uses $D_{\text{NEOWISE}}/\bar{D} = D_{\text{NEOWISE}}/0.5(D_{IRAS} + D_{\text{AKARI}})$. The three rows for "Usui method, uncorrelated" in Table 2 show the results of amending the method of Usui et al. 2014 in this way to remove spurious correlations. Applying all of the adjustments described above results in a 68.27% CI for NEOWISE of −13.6% to +15.9%.

Rather than make a three-way comparison of NEOWISE, *IRAS*, and AKARI diameter estimates, the analysis in M2018b instead performed pairwise comparisons. Comparing the fully cryogenic mission of NEOWISE to AKARI, M2018b found a 68.27% CI of −15.5% to +16.0% for the 3998 asteroids common to those two diameter data sets (Table 13 of M2018b). The CI for the comparison of NEOWISE to *IRAS* was −16.3% to +17.1% (Table 10 of M2018b). These findings are consistent with the expectation that $D_{\text{NEOWISE}}/0.5(D_{IRAS} + D_{AKARI})$ will have less scatter than a comparison to *IRAS* or AKARI alone.

| Source | $D_{\text{NEOWISE}}/\bar{D}$ | | | $D_{IRAS}/\bar{D}$ | | | $D_{AKARI}/\bar{D}$ | | |
|---|---|---|---|---|---|---|---|---|---|
| | **Median** | **68.27% CI** | | **Median** | **68.27% CI** | | **Median** | **68.27% CI** | |
| Usui et al. 2014, Table 2 | 1.006 | 0.913 | 1.099 | 0.992 | 0.898 | 1.086 | 1.001 | 0.925 | 1.077 |
| Usui method, this study | 1.012 | 0.927 | 1.081 | 0.983 | 0.918 | 1.059 | 1.002 | 0.940 | 1.065 |
| + uncertainty | 1.011 | 0.912 | 1.094 | 0.985 | 0.891 | 1.087 | 1.002 | 0.929 | 1.079 |
| + don't average NEOWISE first | 1.007 | 0.902 | 1.103 | 0.985 | 0.889 | 1.091 | 1.002 | 0.926 | 1.084 |
| Usui method, uncorrelated | 1.019 | 0.894 | 1.126 | 0.975 | 0.882 | 1.092 | 1.003 | 0.913 | 1.101 |
| + uncertainty | 1.017 | 0.879 | 1.143 | 0.977 | 0.862 | 1.114 | 1.003 | 0.902 | 1.118 |
| + don't average NEOWISE first | 1.010 | 0.864 | 1.159 | 0.978 | 0.858 | 1.120 | 1.003 | 0.897 | 1.125 |

**Table 2. Replicating the comparison of Usui et al. 2014 with variations.** Top row: distribution statistics from Usui et al. 2014 for ratios of diameter estimates made for 1993 asteroids in the NEOWISE, *IRAS*, and AKARI studies. Median and 68.27% confidence interval were computed from the parameters of the normal distribution that best fit a histogram of the ratios, as described in section 3.1. Usui method, this study: median and CI computed directly from the ratios, rather than from the best-fit normal distribution. +uncertainty: statistics when distribution includes uncertainties in the diameter estimates obtained by using Monte Carlo methods described in section 3.1. +don't average NEOWISE: both uncertainties and all diameter estimates included for asteroids having multiple estimates in the NEOWISE data set. Usui method, uncorrelated: each estimate was compared to the mean of the estimates from the other two surveys, rather than to the mean estimate from all three surveys.

## 3.2     *Distributions of diameter ratios*

W2018, Usui et al. 2014, and M2018b all agree on a general approach for determining overall accuracy: use the ratio of diameter estimates for asteroids where we have a comparison in order to estimate the general distribution of such ratios across the entire population of NEOWISE asteroids.

W2018 Figures 1 and 6 and their supporting text and captions perform this analysis by calculating the simple diameter ratio $D_{\text{NEOWISE}}/D_{\text{ROS}}$ for each asteroid in the population sample. The final comparison numbers are derived by using descriptive statistics (medians, quantiles, or parameters of best-fit normal distributions) for the sets of ratios. Usui et al. fit normal distributions. W2018 appears in places to do the same in some places, but in others they compute quantiles.

Regardless of which descriptive statistics are used, the statistical concept underlying the methods of W2018 and Usui et al. 2014 is to estimate the sampling distribution $\mathcal{R}(\rho)$ of finding a diameter ratio $\rho$ across the population of asteroids under comparison. In other words, given an asteroid at random, what is the likelihood of finding ratio $\rho$?

Having done this for the sample under consideration (for example, the asteroids which have both NEOWISE and ROS diameter estimates), the tacit assumption is that this can be generalized more broadly to all asteroids in the NEOWISE data set. This much is agreed by M2018b with W2018 and Usui et al. 2014; the disagreement is over how to compute the distribution, and in particular whether it is permissible to ignore uncertainties in diameter estimates.

In using the simple ratio of the $D_{\text{NEOWISE}}/D_{\text{ROS}}$, W2018 implicitly assumes that both the NEOWISE and ROS estimates are completely certain, i.e., that they have *zero variance*. In actuality, each set of diameter estimates has an associated estimate of standard error. In the case of NEOWISE, the error estimate comes from propagating the measurement errors in the observations, and errors in other relevant parameters, through the thermal model via a Monte Carlo model. Expressed formally, the NEOWISE diameter estimate is represented by normally distributed random variables $X \sim \mathcal{N}(D_{\text{NEOWISE}}, \sigma_{\text{NEOWISE}})$. Similarly, ROS diameter estimates have propagated the measurement errors relevant to the ROS observations forward to produce an uncertainty, which may be modeled as $Y \sim \mathcal{N}(D_{\text{ROS}}, \sigma_{\text{ROS}})$. Here, the diameter estimates $D$ are the means, and the $\sigma$ are the estimated uncertainties.

The ratio of two random variables is known in statistics as a ratio distribution (Bonamente, 2013; Stuart and Ord, 2010). The ratio distribution of two normal (Gaussian) distributions is *not* a simple normal distribution; it can be quite complicated. Although analytical formulas describe the ratio distribution of two normal distributions (Hinkley, 1969; Korhonen and Narula, 1989; Marsaglia, 1965; Pham-Gia et al., 2006), they have infinite mean and variance because the normal distribution $\mathcal{N}(\mu, \sigma)$ is defined from $-\infty$ to $\infty$. That interval includes zero, and the probability of that occurring, however small, causes the expressions for the mean and variance to diverge. This aspect of the formulas limits their practical utility. When we model the uncertainty of an asteroid diameter estimate $D$ with the normal distribution, there is an implicit assumption that $D \leq 0$, which is not captured in the exact formulas.

While the typical formalism of the normal distribution fails to be of use, one can easily show that the mean of the ratio $X/Y$ is given by the ratio of the means of $X$ and $Y$, or in our case by $D_{\text{NEOWISE}}/D_{\text{ROS}}$ distribution (Bonamente, 2013; Stuart and Ord, 2010). The variance of the ratio distribution is then approximately given by

$$\text{Var}\left(\frac{X}{Y}\right) \approx \left(\frac{D_{\text{NEOWISE}}}{D_{\text{ROS}}}\right)^2 \left(\frac{\sigma_{\text{NEOWISE}}^2}{D_{\text{NEOWISE}}^2} - \frac{2\,\text{Cov}(X,Y)}{D_{\text{NEOWISE}} D_{\text{ROS}}} + \frac{\sigma_{\text{ROS}}^2}{D_{\text{ROS}}^2}\right), \tag{1}$$

where $\text{Cov}(X, Y)$ is the covariance between the estimates.

The approximation to the variance in the ratio of Equation (1) is useful primarily for providing an analytical context. In practice, one can use Monte Carlo trials to sample each distribution. That method, which is used in M2018b, avoids the zero and negative values of $D$ that cause problems for the exact analytical formula.

To understand the impact of Equation (1), consider an example. Suppose an asteroid for which $D_{\text{NEOWISE}} = 100 \pm 10$ km and $D_{\text{ROS}} = 100 \pm 10$ km—i.e., identical estimates and uncertainties. Can we then infer that the NEOWISE estimate is exact?

The approach of W2018 would say yes, it is exact, and record that asteroid as having a diameter ratio of *exactly* 1.0 with zero variance. Using Equation (1), however, we can see that while the mean of the ratio distribution is 1.0, it does *not* have zero variance. If we assume that the distributions are independent, and thus $\text{Cov}(X, Y) = 0$, then the standard deviation of the ratio distribution is approximately 0.141, reflecting contributions from both $\sigma_{\text{NEOWISE}}$ and $\sigma_{\text{ROS}}$. This uncertainty in the *ratio* is equivalent to a 14 km uncertainly in diameter, a factor of $\sim\sqrt{2}$ greater than the original uncertainty. The ratio in this example would be approximately distributed as 1.0±0.14.

The example illustrates that the notation 100±10 is mathematical shorthand for a bundle of estimates—i.e., it is a random variable. When we take the ratio of two of independent random variables, the ratio is also a random variable that has its own variance, approximated by Equation (1).

Consider now a second example of three asteroids having a plausibly realistic range of uncertainties: the first asteroid as in the example above, a second having $D_{\text{NEOWISE}} = 100 \pm 1$ km and $D_{\text{ROS}} = 100 \pm 1$ km, and a third that has $D_{\text{NEOWISE}} = 100 \pm 24$ km and $D_{\text{ROS}} = 100 \pm 10$ km. In all three cases, the mean value of diameter ratios will be 1.0, but the uncertainties of the ratio distributions will differ considerably for each object. The approach used in W2018 ignores the uncertainty in the ratio distribution, so it would treat these three cases as equivalent. Clearly, however, the scatter in the ratio distribution for asteroids 1 and 2, which have 1% uncertainty in both NEOWISE and ROS diameter estimates, must differ from the scatter for asteroid 3, where the estimates have 24 times greater uncertainty.

NEOWISE determined the diameter uncertainty estimates via Monte Carlo analysis that performed model fitting under the assumption of variations in observed and assumed parameters. The number of *WISE* observation data points varies among asteroids, and the uncertainty in NEOWISE diameter estimates thus varies considerably. About 95% of asteroids in the NEOWISE data set have diameter uncertainty in the range ±0.076% to ±32%, a 95% CI that spans a factor of 42. The 68.27% CI for NEOWISE uncertainty is ±2.37% to ±24.7%, a range that spans more than an order of magnitude.

Corresponding ranges for other major surveys are somewhat smaller. The 95% CI in uncertainty for *IRAS* spans a factor of 6.7; for AKARI, a factor of 8.6. ROS uncertainty estimates also vary considerably, from spacecraft visits that are highly accurate to other methods that have considerable uncertainty. These figures show that diameter estimates can vary widely in their estimated uncertainty, which only underscores the importance of using that uncertainty in the accuracy analysis.

The ratio distribution is widely used in many parts of science, including astronomy (González-Marcos et al., 2016; Israel, H. et al., 2017; Mandel et al., 2019; Melchior and Viola, 2012), ecology (Harrison, 2011; Hedges, L. V., Gurevitch, J. and Curtis, 1999), electrical engineering (Mekić et al., 2012; Swamy et al., 2012), economics (Arbutina and Kovačević, 2019; Ezzamel et al., 1987; Ly et al., 2019), food science (Dhanoa et al., 2018), geophysics (Kordas and Petrakos, 2017), materials research (Clemon and Zohdi, 2018), and molecular biology (Andrec et al., 2005; Thomaseth and Radde, 2016). In all these applications, it is important to model the variance in the ratio distribution, rather than to ignore it, as advocated by W2018.

Ignoring the uncertainty in each diameter estimate is also out of keeping with the general principle in quantitative science that we must propagate errors through our analysis so that we estimate the final uncertainty in the results (Dieck, 2017; Mandel, 1964; Taylor, 1997). Indeed, there is little point in estimating the uncertainty in our diameter estimates if we then fail to use that uncertainty when estimating accuracy. Thus, to understand the ratio of NEOWISE and ROS diameter estimates for even a single asteroid, we must treat the ratio as a ratio distribution, which accounts for the uncertainties of the diameter estimates in both the numerator and denominator.

To calculate the ratio for a population of asteroids (e.g., those for which ROS estimates are available), we should also calculate the ratio distribution for each of them. In statistical terms, a distribution of a set of equally weighted, individual distributions is known as a mixture distribution (Frühwirth-Schnatter, 2006; McLachlan and Peel, 2000). The sampling distribution of ratios across asteroids $\mathcal{R}(\rho)$ is the mixture distribution of the ratio distributions for each asteroid. This is entirely consistent with the interpretation of $\mathcal{R}(\rho)$ as estimating the probability distribution for which an asteroid chosen at random will have a given ratio of NEOWISE diameter to ROS diameter.

In compiling ratio samples to estimate $\mathcal{R}(\rho)$ for the population of asteroids, W2018 (and also Usui et al. 2014) use only one sample per asteroid. That is equivalent to approximating the ratio distribution for each asteroid by a single value at its mean. Because the actual variance is non-zero, the ratio distribution is poorly approximated by assuming that it is equal to its mean, as illustrated by Figure 3.

The correct way to estimate $\mathcal{R}(\rho)$ is to use multiple samples. Figure 6(d) of W2018 plots a histogram of the ratios of $(D_{\text{NEOWISE}} - D_{\textit{IRAS}}) / D_{\text{NEOWISE}}$ for ~1700 asteroid diameters from *IRAS* and NEOWISE, using 2018 code. W2018 reports that a normal distribution fit to the histogram has $\sigma = 0.0991$.

The impact of Equation (1) in this case can be seen by assuming, for example, ±10% uncertainty for each NEOWISE and *IRAS* diameter estimate. From Equation (1), the uncertainty in the ratio is ±14%, which must be applied to each of the diameter ratios that make up the histogram. To estimate the impact, we can use the formula for the sum of two independent normally distributed random variables $\sigma_{1+2} = \sqrt{\sigma_1^2 + \sigma_2^2}$, where $\sigma_1, \sigma_2$ are the uncertainties of each. Under these assumptions, the scatter depicted in Figure 6(d) accounting for uncertainty in NEOWISE and *IRAS* estimates would be $\sigma \approx \sqrt{0.0991^2 + 0.14^2} \approx 0.17$, or $\pm 17\%$, rather than $\pm 9.91\%$ as indicated in W2018. Applying the same process to Figure 6(e), which plots comparable ratios for NEOWISE 2011, we obtain $\sigma \approx 0.19$ or $\pm 19\%$.

Note that while these figures illustrate the impact of the ratio distribution variance, they are idealized. In actuality, the uncertainties vary for each NEOWISE and *IRAS* diameter estimate rather than being a uniform 10%. In addition, the caption and supporting text for Figure 6 do not state how the comparison between *IRAS* and NEOWISE treats the 117 asteroids having copied ROS diameters, ~87 of which are in the set. Finally, the NEOWISE estimates for Figure 6 presumably have been averaged when there are more than one (section 3.3).

Rather than attempting to use Equation (1), M2018b demonstrated an improved method that approximates the ratio distribution by using Monte Carlo methods to create multiple samples for the numerator and denominator distributions, using the correct uncertainty for each, and then takes their ratio to find samples of the ratio distribution. As before, the samples are compiled into a population sample, which is then characterized by descriptive statistics or by fitting a parametric distribution. M2018b compared diameter estimates for 1726 asteroids computed by NEOWISE from the 2011 papers to estimates published from *IRAS* and found that the ratio distribution has a median of 1.024 (i.e. +2.4%), with a 68.27% CI of -16.3% to +17.1% (see Table 10 in M2018b). For the 87 cases in which the 2011 NEOWISE diameter estimates were actually copied ROS diameters, the median is 0.993 (-0.7%), with a 68.27% CI of -13.4% to +12.2%. Removing those cases increases the scatter between NEOWISE and *IRAS* substantially, to a median of 0.635 (-36.5%), with a 68.27% CI of -35.9% to +113.4%.

Figure 3 uses the Monte Carlo method to illustrate the effect of uncertainty on the ratio distribution. The process in this illustrative example begins by assuming an asteroid of true diameter 100 km. A Monte Carlo code was used to draw 10,000 simulated NEOWISE samples having normally distributed error of $\pm 8\%$ from the true diameter. The same number of simulated ROS samples were then drawn; these had a normally distributed error of $\pm 6\%$. Note that the results in the figure and described below are independent of the actual true diameter used. As seen in the top left panel of Figure 3, the example histogram of simple ratios of simulated NEOWISE and ROS diameters is slightly asymmetric, as was seen in Usui et al. 2014.

Because each NEOWISE and ROS diameter estimate has an associated uncertainty of $\pm 8\%$ and $\pm 6\%$, respectively, each NEOWISE estimate is a normally distributed random variable whose mean is given by the diameter drawn in the first round of the simulation, with uncertainty $\pm 8\%$. One example pair having $D_{\text{NEOWISE}} = 91.9 \pm 7.35$ km and $D_{\text{ROS}} = 112.5 \pm 6.75$ km was selected and used to draw 10,000 samples from those distributions (Figure 3, top center). Dividing these samples yields a histogram that represents the ratio distribution for this example synthetic asteroid (Figure 3, top right).

In the method advocated by W2018, this particular asteroid pair would be represented only by the ratio of the means: $91.9/112.5 \approx 0.817$, ignoring the uncertainty in each estimate.

A corresponding ratio distribution can be calculated for each pair of corresponding NEOWISE and ROS estimates. As seen in the 5 examples pictured here (Figure 3, bottom left), the means and confidence intervals of these distributions can vary considerably. Combining 10,000 samples for each of the 10,000 pairs yields a distribution of $10^8$ samples (Figure 3, bottom center) that expresses $\mathcal{R}(\rho)$, the probability of finding a given ratio in this population of simulated asteroids.

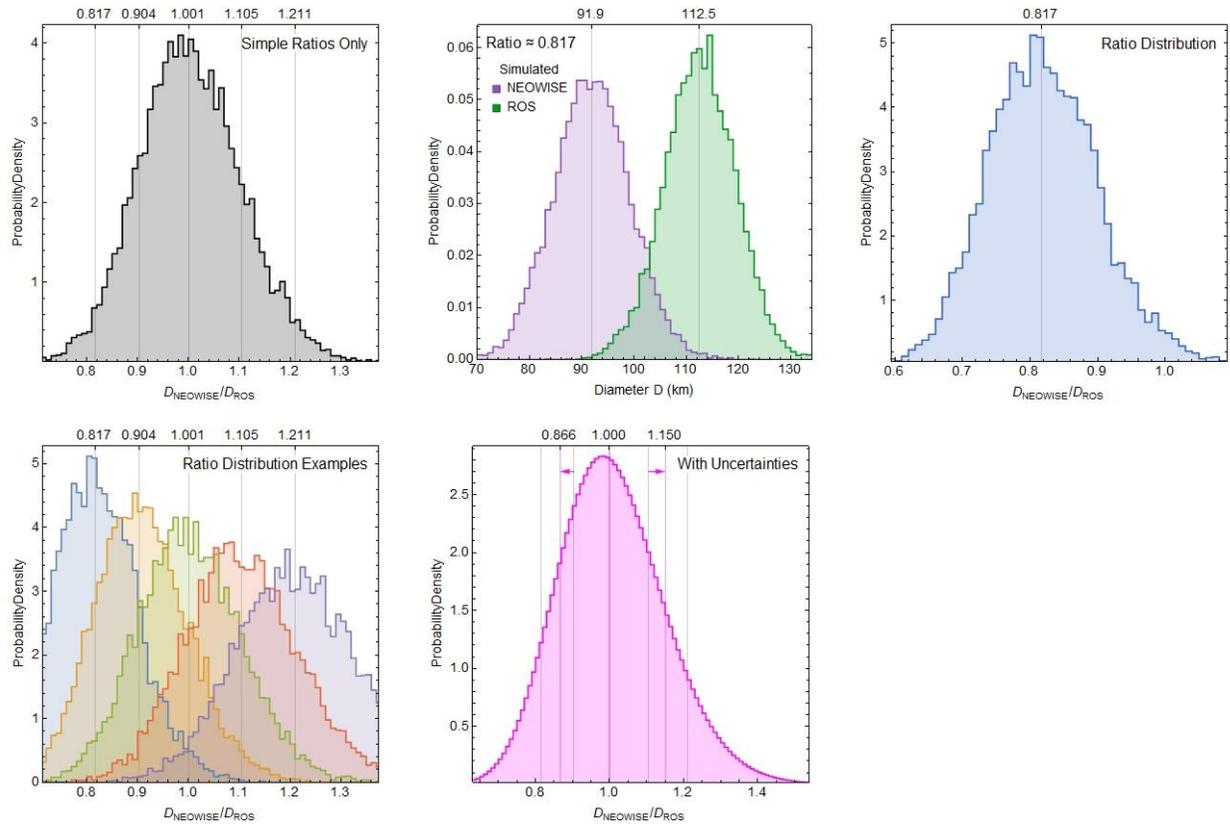

**Figure 3. Monte Carlo histograms of the effect of uncertainty on the ratio distribution.** To simulate NEOWISE results, 10,000 Monte Carlo samples were drawn, using a true asteroid diameter of 100 km and a normally distributed standard error of $\pm 8\%$; the same number of samples were drawn to simulate ROS estimates with SE of $\pm 6\%$. Each simulated NEOWISE and ROS estimate is assumed to have the same corresponding uncertainty. Top left: histogram of the simple ratios of these samples, without their uncertainty. Vertical lines indicate the median ratio and the 68.27% and 95% confidence intervals. By choosing one of these ratios we can draw samples to illustrate the uncertainty in each. Top center: Histograms of NEOWISE and ROS diameters for 10,000 Monte Carlo samples drawn using the example ratio $D_{\text{NEOWISE}} = 91.9 \pm 7.35$ km and $D_{\text{ROS}} = 112.5 \pm 6.75$ km. The ratio of the mean of each distribution is 0.817. Top right: A histogram generated by sampling the ratio distribution of these two random variables and dividing the samples shows the uncertainty in the ratio. The same can be done for each NEOWISE, ROS pair. Bottom left: 5 example histograms of 10,000 samples each. Bottom center: Accumulating the 10,000 ratio distribution samples for each NEOWISE, ROS simulated pair yields a histogram over $10^8$ samples. The effect of taking into account the uncertainty in ratio given by the ratio distribution is to increase the 68.27% confidence interval for the simple ratios from 0.904 to 1.105 (*top left*) and from 0.866 to 1.15 (*bottom center*) when accounting for uncertainty (*magenta arrows*).

### 3.3   Handling multiple estimates for the same asteroid

NEOWISE produced a single diameter estimate for most of the asteroids it studied, but in less than 10% of cases, several independent NEOWISE diameter estimates (up to 4 for some asteroids) were created from different sets of observations, even within the same mission phase (i.e., the fully cryogenic mission phase). Multiple ROS diameter estimates are similarly available for only a small subset of those asteroids that have any ROS estimate.

A specific example is shown in Figure 4, which was Figure S9 in M2018b. The left panel presents the multiple NEOWISE and ROS estimates for asteroid 192 Nausikaa, for which NEOWISE has published 4 diameter estimates, while three additional estimates are available from ROS papers. As noted in section 3.2, each estimate has an associated—and in this example, different—uncertainty associated with it.

W2018 advocates averaging all available estimates from NEOWISE or ROS sources, respectively, prior to comparing them—i.e. computing $\bar{D}_{\text{NEOWISE}}/\bar{D}_{\text{ROS}}$ for each asteroid. This raises the question of the goal of the comparison exercise. If the goal is to characterize the very best cases where NEOWISE has multiple estimates, then the averaging-before-comparing method would be valid. However, if our goal is to estimate the overall accuracy of NEOWISE estimates, then this is not appropriate. An accuracy metric based on $\bar{D}_{\text{NEOWISE}}/\bar{D}_{\text{ROS}}$ is reasonable when limited to the 10% of asteroids for which multiple estimates are available. But that metric is not applicable to the 90% of asteroids in the NEOWISE data set for which only a single diameter estimate has been made.

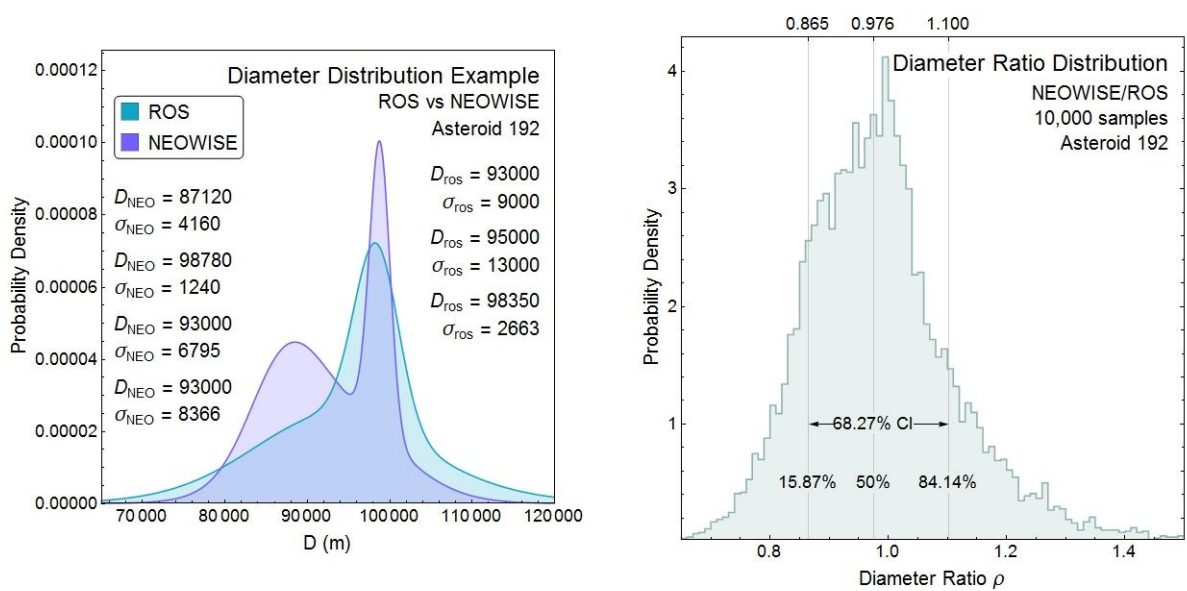

**Figure 4 (reproduced from Figure S9 of M2018b).** Statistical evaluation of diameter estimation accuracy is illustrated for asteroid 192 Nausikaa. Left: Multiples values of diameter *D* estimated from NEOWISE thermal modeling and from radar and other observations, which are known from their statistical distributions (M2018b). Right: Monte Carlo samples drawn from these distributions were divided to produce a histogram of the diameter ratios. Vertical lines mark the median and bounds (at 15.87% and 84.14% quantiles) of the 68.27% (~1σ) confidence interval.

M2018b treats multiple estimates as a mixture distribution, as described in section 3.2. In the example of asteroid 192 (Figure 4), the multiple NEOWISE and ROS estimates on each side allow 12 pairwise comparisons. One could use Monte Carlo methods to sample the ratio distribution for each of these pairs and then weight each ratio by 1/12, a process that can be considered as averaging the samples *after* ratio calculation, in contrast to the approach advocated in W2018 of averaging estimates *before* the ratio is calculated.

In practice, it is simpler and more efficient to draw trials directly from the mixture distributions. Doing this for the example of asteroid 192 yields a histogram of the resulting samples of the ratio distribution (Figure 4, right panel). This approach produces generalizable results because the same overall weighting to samples is obtained for an asteroid having 12 diameter pairs as occurs for an asteroid that has only one pair.

The W2018 approach of averaging before comparison would be a legitimate way to determine the best pooled NEOWISE estimate and to compare it to the best ROS estimate pooled across multiple ROS estimates. In each case, averaging the multiple estimates will reduce the scatter present in a single estimate. However, W2018 presents the results obtained from average-first metric as a *typical* accuracy value that characterizes other cases. That metric actually characterizes *only* their very best cases.

The W2018 objection to the average-after-comparison method of M2018b hinges on this question: what is the goal of the accuracy estimation? W2018 advocates a method that tries to find the very best estimates and then take their ratio. In contrast, M2018b attempts to find a method that yields a result that is valid for typical cases.

As an example, across the NEOWISE papers covering the fully cryogenic mission, 9.6% of the asteroids have more than one NEOWISE diameter estimate. If we look at the subset of 1728 asteroids that these papers have in common with *IRAS*, 71% have more than one NEOWISE diameter estimate. That is about 7.3 times as many of that kind as are found in the population as a whole. Averaging before comparison thus unjustifiably boosts apparent accuracy by a significant amount for the subset in common with *IRAS*.

W2018 mischaracterizes the mixture-distribution approach—a standard statistical technique—as "adding their probability distributions instead of multiplying them." A mixture distribution is *renormalized*, not just a sum. It is in fact identical to the method that W2018 uses, except that the calculation in W2018 is the mixture distribution of exact ratios. W2018 assumes, in other words, that each ratio is exact. As discussed in section 3.2 above, this assumption is incorrect. Under no reasonable interpretation of this situation would it be appropriate to multiply the probability distributions, as stated in the quote.

## 3.4   The W2018 Monte Carlo model

As a further argument in favor of its approach, Figure 2 of W2018 presents results from a Monte Carlo model created from a set of synthetic ROS and NEOWISE data estimates based on 50 hypothetical "true" diameters $D_{\text{True}}$. These diameters are related to the estimates under the assumption that they obey the following error model:

$$\begin{aligned}
\widehat{D}_{\text{ROS}} &= D_{\text{True}} + c & c &\sim \mathcal{N}(0, \sigma_{\text{ROS}}) & \sigma_{\text{ROS}} &= 0.12\, D_{\text{True}} \\
\widehat{D}_{\text{NEOWISE}} &= D_{\text{True}} + d & d &\sim \mathcal{N}(0, \sigma_{\text{NEOWISE}}) & \sigma_{\text{NEOWISE}} &= 0.08\, D_{\text{True}} \\
D_{\text{ROS}} &= \widehat{D}_{\text{ROS}} + e, & e &\sim \mathcal{N}(0, \sigma_{\text{ROS}}), & \sigma_{\text{ROS}} &= 0.12\, D_{\text{True}} \\
D_{\text{NEOWISE}} &= \widehat{D}_{\text{NEOWISE}} + f, & f &\sim \mathcal{N}(0, \sigma_{\text{NEOWISE}}), & \sigma_{\text{NEOWISE}} &= 0.08\, D_{\text{True}}. \\
D_{\overline{\text{ROS}}} &= \tfrac{1}{4}(\widehat{D}_{\text{ROS}_1} + \widehat{D}_{\text{ROS}_2} + \widehat{D}_{\text{ROS}_3} + \widehat{D}_{\text{ROS}_4}), \\
\overline{D_{\text{ROS}}} &= D_{\overline{\text{ROS}}} + g, & g &\sim \mathcal{N}(0, \sigma_{\text{ROS}}/2)
\end{aligned} \quad (2)$$

The 50 values of $D_{\text{True}}$ are uniformly distributed in log space between 20 and 500 km, although the results are independent of $D_{\text{True}}$ because a ratio is taken of $D_{\text{NEOWISE}}/D_{\text{ROS}}$. For each $D_{\text{True}}$, 4 synthetic ROS diameter estimates $\widehat{D}_{\text{ROS}}$ are randomly drawn from a normal distribution having the specified $\sigma_{\text{ROS}}$. One synthetic NEOWISE estimate $\widehat{D}_{\text{NEOWISE}}$ is made by drawing from a normal distribution of specified $\sigma_{\text{NEOWISE}}$. W2018 further specifies that the uncertainty of each synthetic diameter estimate $D_{\text{ROS}}$, $D_{\text{NEOWISE}}$ is given by the same $\sigma_{\text{ROS}}$, $\sigma_{\text{NEOWISE}}$. The 4 synthetic ROS estimates are averaged to get $D_{\overline{\text{ROS}}}$, which has uncertainty $\sigma_{\text{ROS}}$.

The result of this procedure is that each asteroid is represented by two normally distributed random variables, $D_\text{NEOWISE}$, and $\overline{D_\text{ROS}}$, each having different means and the specified uncertainty.

W2018 uses this framework to create a histogram of $\widehat{D}_\text{NEOWISE}/D_{\overline{\text{ROS}}}$ (right panel of Figure 2 of W2018), from which they claim the "correct" 16th and 84th percentiles (approximately the 68.27% CI) of 0.88 and 1.08.[2]

Generating 4 ROS estimates with $\sigma_\text{ROS} = 0.12\, D_\text{True}$, only to immediately average them, yields a mathematically identical result to drawing a single ROS estimate with $\sigma_\text{ROS} = 0.06\, D_\text{True}$. This example is therefore applicable only to a case where *every* asteroid has 4 independent ROS estimates of ±12% accuracy versus the true diameter, or equally to a case where each asteroid has one ROS estimate at ±6%. However, it is fallacious to apply this example to the NEOWISE data set, in which most asteroids have one ROS estimate at ±12% and 4 estimates are available for only a tiny fraction of asteroids. The method used in M2018b, in contrast, is designed to be applicable to cases having a single ROS estimate.

W2018 treats the simple ratio $\widehat{D}_\text{NEOWISE}/D_{\overline{\text{ROS}}}$ as exact in using it to compile the histogram in their Figure 2. Following the procedure specified in W2018 produces highly variable results; 50 samples are far too few Monte Carlo trials to generate stable 16th and 84th percentiles. Repeating the procedure with the specified 50 $D_\text{True}$ examples 10,000 times, yields 16th and 84th quantiles of 0.904 and 1.103, respectively.

The caption to W2018 Figure 2 states:

> Histograms for the direct comparison of $D_\text{WISE}$ vs. $D_\text{ROS}$ are shown in black; comparisons based on the non-standard M2018b approach are shown in red. The black histogram produces the expected 16th and 84th percentile ratio of $D_\text{WISE}$ vs. $D_\text{ROS}$ of 0.88 and 1.08, respectively, whereas the red histogram produces 16th and 84th percentile ratios of 0.82 and 1.21.

The 16th and 84th percentiles of a normal distribution with standard deviation ±10% are not 0.88 to 1.08. It is unclear whether the error here in W2018 arises in the caption, the calculation, or an omission of a crucial aspect of the procedure.

Although the ±10% variance of the diameter ratio obtained by Monte Carlo analysis in this synthetic example of W2018 does match the value obtained from a standard formula (i.e., $\sqrt{8^2 + 6^2} = 10\%$), that equivalence rests on the assumption that each simulated NEOWISE and ROS estimate is exact.

The correct approach is to take the uncertainty in diameter estimates into account, as M2018b does (section 3.2). Treating synthetic estimates as if they are exact fails to realistically simulate NEOWISE and ROS estimates.

---

[2] For ease of comparison with W2018, I have in this section reported values for 16th and 84th percentiles, as W2018 does. These percentiles approximate, but are not identical to, a standard deviation ($1\sigma$) and are thus harder to compare to standard results. When reporting the results of Monte Carlo simulations, it is preferable to use quantiles because the resulting distributions are not normal distributions and can be asymmetric. The quantiles that most closely approximate the $1\sigma$ standard deviation are 0.15875 and 0.84125, which are the boundaries of the 68.27% confidence interval.

When taking these uncertainties into account in their model and using 10,000 Monte Carlo trials for each synthetic asteroid, the 16$^{th}$ and 84$^{th}$ percentiles are 0.865 to 1.148. If instead we use the $D_{ROS}$ without averaging (hence with twice the uncertainty), then the 16$^{th}$ and 84$^{th}$ percentiles are 0.841 to 1.195. Neither case matches the assertion in the quoted passage about the formula $\sqrt{2(8^2 + 12^2)} = 20.4\%$. That formula does not make sense here, in part because there is no simple formula for the variance of a ratio distribution. Equation (1) is an approximation, but it is not consistent with this calculation either.

In practice, the diameter estimate uncertainty in NEOWISE discussed here reflects within-model *precision*, not model accuracy (Dieck, 2017; Mandel, 1964; Taylor, 1997). A more realistic model for the errors is

$$D_{ROS} = D_{True} + e + \lambda_{ROS}, \quad e \sim \mathcal{N}(0, \sigma_{ROS}), \quad \sigma_{ROS} = 0.12\, D_{True}$$
$$D_{NEOWISE} = D_{True} + f + \lambda_{NEOWISE}, \quad f \sim \mathcal{N}(0, \sigma_{NEOWISE}), \quad \sigma_{NEOWISE} = 0.08\, D_{True} \quad (3)$$

where the random variables *e* and *f* are the measurement or modeling uncertainty, and $\lambda_{ROS}$ and $\lambda_{NEOWISE}$ are unknown random variables that depend on out-of-model effects, some of which are discussed in section 3.5. In standard statistical terms, *e* and *f* are the precision terms—the uncertainty in measurements, as expressed in the modeling. Systematic bias is due to $\lambda_{ROS}$ and $\lambda_{NEOWISE}$, which contribute along with *e* and *f* to random errors in the model.

In terms of Equation (3), the W2018 error model of Equation (2) effectively states that $\lambda_{ROS}$ and $\lambda_{NEOWISE}$ are distributed in the same way as *e* and *f*. While on cursory inspection it may appear that one is double counting the uncertainty, this is not the case, as Equation (3) makes explicit.

Actual NEOWISE and ROS estimates for asteroid 192 Nausikaa are shown in Figure 4 (left panel). The uncertainties $\sigma_{ROS}$, $\sigma_{NEOWISE}$, characterize the statistical distributions of *e* and *f*. The diameter estimates are $D_{True} + \lambda_{ROS}$ and $D_{True} + \lambda_{NEOWISE}$—certainly *not* all the same value $D_{True}$. The NEOWISE papers estimated $\sigma_{NEOWISE}$, the standard error for *f*, by using a Monte Carlo calculation that intended to account for variations in observed quantities (uncertainty in the *WISE* fluxes) and in parameter assumptions (i.e., for the photometric phase-law slope *G*). A similar analysis of measurement and estimation errors applies for ROS estimates; those errors constitute *e*.

### 3.5   Diameter comparison plots

W2018 mischaracterizes two of the figures in M2018b:

M2018b has Figures 7 & 8 that at first glance look similar to Figure 1, but with much larger central 68% confidence intervals. This is caused by two non-standard procedures used by M2018b. The first non-standard procedure is to plot not $D_{ROS}/D_{WISE}$ but rather ratios of Monte Carlo samples drawn from the published means and error estimates.

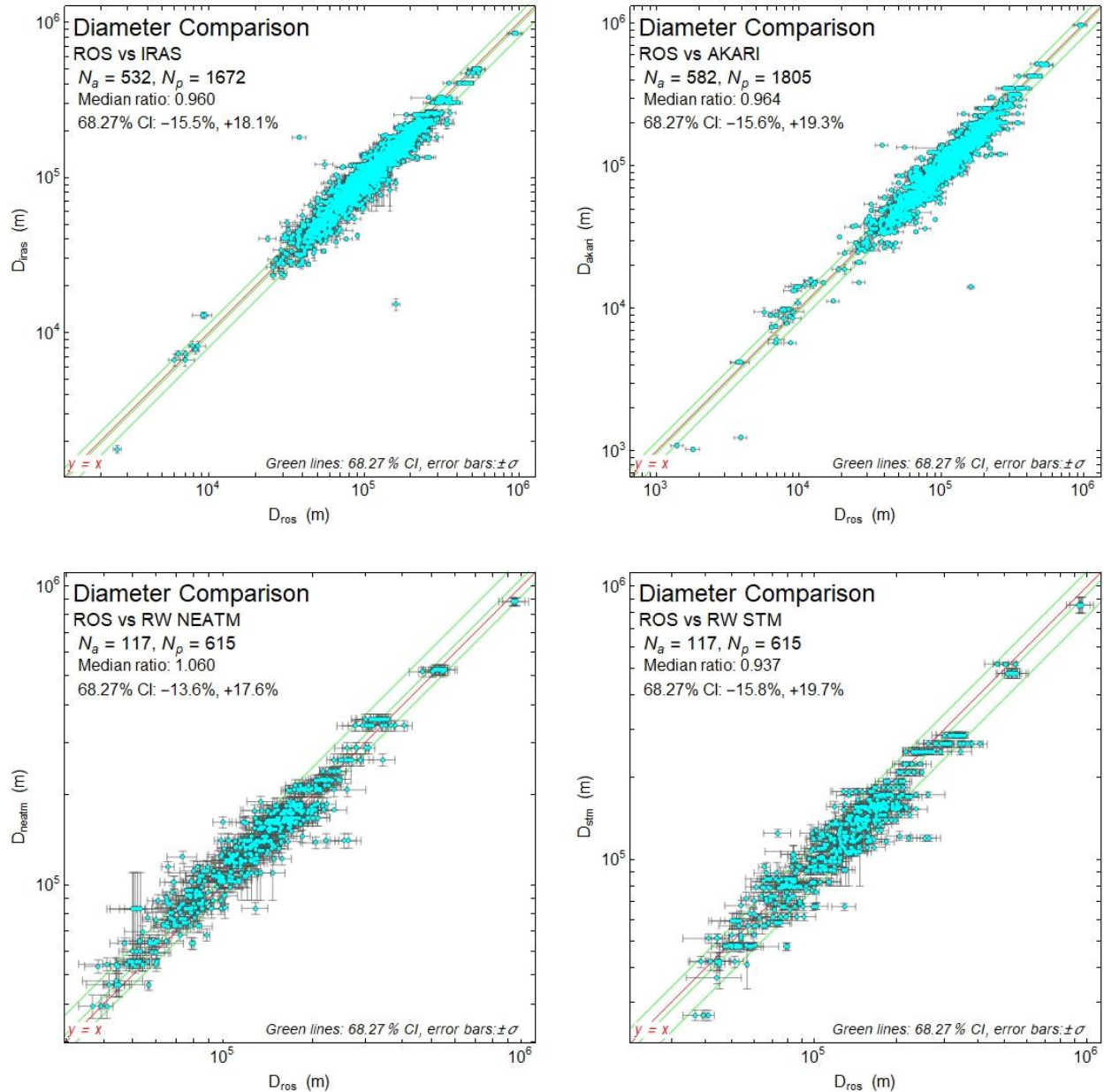

**Figure 5 (reproduced from Figure 7 of M2018b). Comparison of diameter estimates derived from non-NEOWISE thermal modeling to estimates made from radar, occultation, and spacecraft (ROS) observations.** $D_{ROS}$, diameters for asteroids from the ROS literature, with $\pm\sigma$ error estimates. $D_{IRAS}$, diameters from thermal modeling by IRAS (Tedesco et al., 2002). diameters from Ryan and Woodward (2010). $N_a$, the number of asteroids. $N_p$, the number of data points (multiple points for some asteroids).

In M2018b, Figure 7 (reproduced here as Figure 5) compares diameter estimates from ROS to estimates from AKARI, *IRAS*, and Ryan and Woodward (2010), and Figure 8 compares ROS estimates to NEOWISE diameters. Both figures plotted the actual diameter estimates and their error bars, as clearly labeled on the plot axes and repeated in the captions. The error bars are based on the errors for each source that were published in the source papers—i.e., the NEOWISE uncertainty in *D* for NEOWISE results and the ROS, AKARI, or *IRAS* errors for those sources.

These figures in M2018b do differ from Figure 1 of W2018, but not in the manner described. The difference is that M2018b plots all NEOWISE estimates in cases where multiple diameters are available, whereas W2018 advocates averaging first (section 3.3). The figure labels and captions to Figures 7 and 8 in M2018 noted the number of data points and the inclusion of multiple data points for some asteroids.

M2018b included all estimates to avoid statistical bias when analyzing the ~10% of asteroids for which NEOWISE made multiple estimates (section 3.3). Each estimate was treated as a separate NEOWISE result, just as they were presented in the original NEOWISE papers and the PDS archive.

## 3.6 Out-of-model errors

As noted in section 3.4, the model examples used to characterize errors in the W2018 synthetic data example (Equation (2)) neglects the crucial distinction in statistics between *accuracy* and *precision* (Dieck, 2017; Mandel, 1964; Taylor, 1997). W2018 poses a thought experiment that involves two ROS estimates for the diameter of an asteroid, one of $90 \pm 10$ km and a second of $110 \pm 10$ km. Wright et al. claim that the "correct" approach is to average the two estimates to yield $100 \pm 7$ km.

The two hypothetical estimates offered in this example are incompatible in the sense that 90 is two standard deviations from 110. It is thus very unlikely that both estimates would be drawn from the same distribution (~2.3% probability either way). The probability of drawing two samples from the distribution 100±7 km that are ≥10 km apart is only 0.5%.

While those are presented as made-up examples, something similar occurs in real cases as well. In the example of asteroid 192 Nausikaa (Figure 4), the NEOWISE diameter estimates range from 87120±4160 m to 98780±1240 m. In terms of the highest estimate, the lowest estimate is 9.4 standard deviations away; with respect to the lowest estimate, the largest diameter is 2.8 standard deviations, in both cases comparing the center of the distribution. It is clearly very unlikely that these reflect truly random variables drawn from the same normal distribution—with either value of the uncertainty.

Several factors contribute to this situation. Among most important is likely the non-spherical shape of many asteroids, which present varying cross-sectional areas as they rotate, resulting in variable light curves. As seen in Figure 6 (left panel), for example, a subset of the *WISE* observations for asteroid 85839 are well fit by a sinusoidal curve of the form flux $= M + (A/2)\sin(\omega t + \varphi)$, where *M* is the mean magnitude, *A* is the light curve amplitude, $\omega$ is the angular frequency and $\varphi$ is the phase. Band W2 is shown, but the other bands have similar light curve variation.

Under the assumption that the variation in W2 is due to changes in the effective cross sectional area, then the range of variation in effective diameter of that cross section is $\Delta D \propto D(10^{-0.1\,A}, 10^{0.1\,A})$.

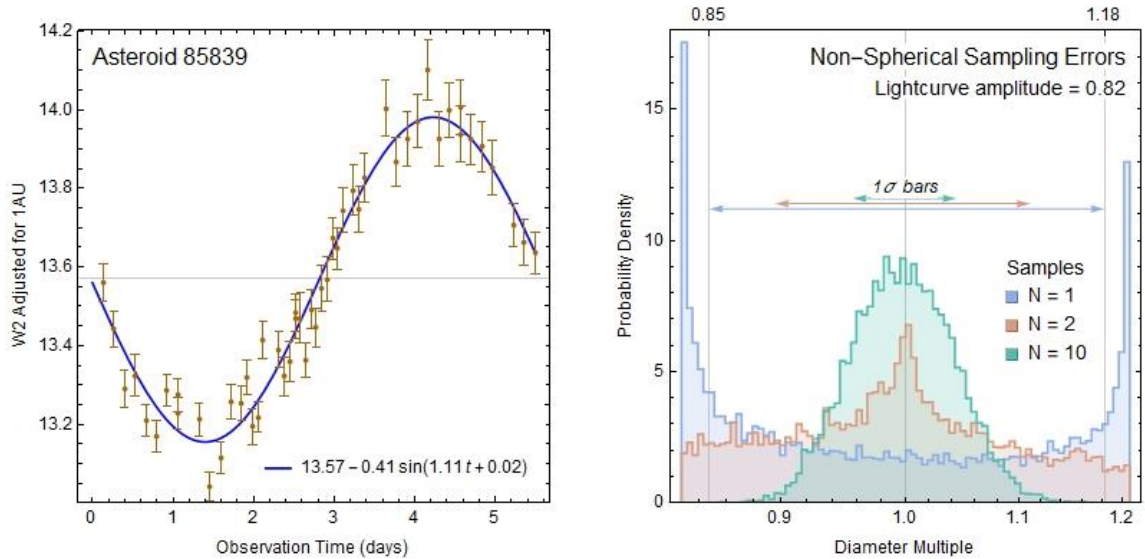

**Figure 6. Effect of the light curve on diameter estimation accuracy.** Left: a subset of the NEOWISE observations in the W2 band for asteroid 85839. Error bars plot measurement uncertainties, as estimated by the *WISE* pipeline. The best fitting sinusoidal curve (*blue*) demonstrates an evident light curve effect, with amplitude 0.82 magnitude. The average of the fitted curve over a full period is 13.57 mag (*horizontal line*). Right: histograms of 10,000 uniform random samples across the light curve period expressed as multiples of the diameter of the effective cross-sectional area. Histograms are shown for single samples ($N = 1$, *blue*), 20,000 samples averaged pairwise ($N = 2$, *orange*), and 100,000 samples averaged in groups of $N = 10$ (*green*). Colored arrows mark the corresponding 68.25% confidence intervals for the three cases. Vertical bars and figures at top mark the bounds of the $N = 1$ confidence interval.

Assuming an asteroid having this light curve, we can estimate the probability of obtaining different diameter estimates in one or more samples of the light curve at randomly chosen rotational phases. Figure 6 plots histograms of such cases. The $N = 1$ case makes a single sample of the light curve; $N > 1$ makes multiple samples at random rotational phase and averages them. A multiple of 1.0 means that the measured diameter equals the diameter of the mean over the light curve period (i.e. the diameter consistent with a flux of $M$). In this example, the $N = 1$ samples (blue) span a range from 0.83 to 1.21 overall, with the most likely points at the two extremes. The 68.27% confidence interval is 0.85 to 1.18.

This histogram shows the uncertainty in any diameter estimate that samples a very short interval on this light curve, even if one has *perfect* measurement. For example, occultation-based diameter estimation involves making an instantaneous estimate of the asteroid's size; one would expect that a single occultation event would thus be similar to the $N = 1$ case. Because actual measurement precision is not perfect, this uncertainty is over and above the estimated uncertainty $\sigma_{ROS}$, which generally does not assume an irregular asteroid shape.

More generally, for $N = 1$, Monte Carlo simulations show that the $1\sigma$ range is $\pm 0.88\,A$ in flux magnitudes, or $\Delta D \propto D\,(10^{-0.088A}, 10^{0.088A})$. Given a typical ROS measurement precision of about 10% (i.e. $\sigma_{ROS} \approx 0.1\,D$), we can estimate the light curve amplitude that produces an error that is important compared to that; for example, a threshold of 2.5% (i.e., a 25% increase compared to typical 10% measurement error for the spherical case) yields $A \geq 0.12$. TALCS (the Thousand Asteroid Light Curve Survey) (Masiero et al., 2009) found $A \geq 0.12$ to be true for roughly 70% of asteroids. It thus represents a significant effect.

    The example here is offered only to illustrate the errors that can be contributed by the light curve effect from non-spherical asteroids. Real light curves are often much more complicated than the simple sinusoidal form found for 85839. In addition, the prevalence of non-spherical shapes is typically assumed to be size-dependent, with more departures from spherical for smaller asteroids. This suggests that the distribution of errors due to inadequate light-curve sampling would be heteroscedastic with respect to diameter.

    Light curve effects also affect some NEOWISE estimates. In contrast to the case of asteroid 85839, where the number of samples is sufficient to average out the light curve, even the highest-quality category of NEOWISE data has a minimum requirement of only three observations per band. In such cases, light curve variations arising from small sample sizes can contribute significant errors to the NEOWISE estimates. The high and low NEOWISE diameter estimates for asteroid 192 Nausikaa (Figure 4) likely reflect unanalyzed errors of this kind.

    Because the NEATM assumes a spherical asteroid, its Monte Carlo error analysis, which uses the NEATM, exclusively computes within-model error—i.e., the extent to which errors in observed inputs impact the model output. A departure from spherical shape is an out-of-model error that is not captured in any of the model inputs. For the same reason, ROS measurements typically offer only within-model estimates that do not include light curve effects.

    Out-of-model errors of a different kind also occur in occultation measurements, which measure chords along the shadow cast on the observers. One rarely knows the longest chord possible. Even larger errors can occur if the occultation captures only one member of a binary asteroid system.

    Out-of-model errors of variously kinds collectively contribute to $\lambda_{ROS}$ and $\lambda_{NEOWISE}$ in Equation (3). W2018 seems to agree that out-of-model errors can be important; light curve error is one part of their explanation for not reporting a critical software bug (section 4.1 below). Elsewhere W2018 mentions modeling errors due to observation at high phase angles, another valid source of error, though one of uncertain importance for NEOWISE, which observes in a highly constrained range of phase angles.

    Although W2018 notes these out-of-model errors, it does not take them into account when it repeats the NEOWISE claim of "$\pm 10\%$" accuracy figures for its diameter estimates. W2018 thus claims that the effects are large enough to figure in their exculpatory explanation for not reporting the software bug, yet not large enough to impact their accuracy claims. A complete error analysis for NEOWISE would remedy this and include all relevant out-of-model errors, which would increase the uncertainty of the NEOWISE diameter estimates.

# 4. Model fits that don't fit the data

*4.1 The NEOWISE software bug*

M2018b offered empirical analysis of the NEOWISE corpus of results because those results have been widely used in the community yet insufficiently documented. As part of that empirical analysis, M2018b discovered that many published NEOWISE curve fits were of poor quality. Among the model cases based on all 4 *WISE* bands with full thermal modeling, for example, 43% to 49% of model fits published by NEOWISE *completely miss* one or more entire bands of data (M2018b section 6 and Figures 3, 4, and S8).

W2018 partially explains this failure by reporting, for the first time, that a bug was discovered and repaired in 2011, evidently sometime after the publication of ~130,000 results in Masiero et al. (2011) and Mainzer et al. (2011c). This bug apparently introduced errors into a large fraction of the NEOWISE results.

W2018 does not explain why the research team did not fulfill its obligation to immediately report errata describing this pervasive error to the published results and to publish corrected results as soon as possible. As a consequence, many of the follow-on studies that relied on the NEOWISE results as a foundation for further work may have been corrupted. The failure by such a prominent group to communicate and correct these errors in a foundational study risks undermining trust in shared databases that are essential to progress in astronomy.

Though it was aware of this bug, the NEOWISE team forcefully resisted and publicly denigrated the independent evaluation in M2018b that described evidence of such a flaw. Any of the many papers that the group has published since 2011 could have explained the error and corrected the record by identifying those previous results they knew to be flawed. None did so.

Examples of NEOWISE fit curves missing the data to be fit were shown to the NEOWISE group in mid–2015 when early drafts of Myhrvold (2018a) and M2018b were shared with the project. In mid-2016, the NEOWISE group, with Joseph Masiero as lead author, deposited NEOWISE results in the Planetary Data System (Mainzer et al., 2016). The supporting text describing the results, and the per-result data labels, repeatedly asserted that the results were *identical* to previously published results. As M2018b documented in detail, this assertion was false. The data archived in the PDS contained thousands of changes from the NEOWISE results originally published in *The Astrophysical Journal*, including the addition of asteroids that never appeared in the original papers, the omission of asteroids that had previously been included, and altered diameters and other results for many objects (M2018b Tables 2 and 3 and section 4.1).

W2018 now acknowledges the disparity as "some minor issues with consistency between tables due to clerical errors." However, independent examination of the changes revealed that while some were minor, most were not and included the addition or deletion of all records for some asteroids (M2018b).

As described in section 2.2, the NEOWISE team has given shifting explanations for the intermingling of copied ROS diameters with modeled diameters in Masiero et al. 2011. M2018b examined the 2016 submission of NEOWISE results to the PDS archive and found that, in *every* instance where Masiero et al. 2011 had included a copied ROS diameter, the PDS entry had been altered to a new diameter (M2018b). M2018b also reported that each of the data records for these modified results had been mislabeled as a model fit published in the original paper by Masiero et al. (2011).

The PDS submission was made with Masiero as lead author at approximately the same time as Masiero was defending the copied values as legitimate in the Yahoo minor planets forum (see quote in section 2.2). So at the about same time he was defending the use of copied diameters, those results were being deleted from the corpus submitted to the PDS and in many cases substituted with results that were backdated. Masiero is also a co-author of W2018, so it seems reasonable to ask that this be explained.

The 2016 NEOWISE submission of results to the PDS offered the group an opportunity to correct diameter estimates that had been corrupted by the 2011 bug. However, the NEOWISE result parameters that W2018 identifies for asteroids 25916 and 90367 as being affected by the bug are identical to the corresponding values entered into the PDS archive in 2016. In addition, M2018b analyzed NEOWISE results as presented in the PDS archive and found that 49% of NEOWISE fits completely missed at least one band of data.

We must therefore conclude that, although thousands of results were altered without explanation or apparent justification in the 2016 PDS submission, results corrupted by the bug were *not* corrected. Receipt of the early draft of M2018b a year earlier was apparently not sufficient reminder to address the corrupted results with a correction. The existence of bug-corrupted results was made public until the final publication of M2018b prompted the release of a preprint of W2018 in November 2018.

W2018 does not dispute this. Regarding the team's failure to report the bug and its consequences, W2018 acknowledges:

> Because the effect of the issue in general is smaller than e.g. the effects of incomplete coverage of lightcurve amplitudes, the team was more focused on quantifying the effects of the lightcurve sampling on the derived diameters, and description of the issue was not published after it was remedied; this was an oversight.

Wright et al. do not provide here tables of pre- and post-correction results to support their assertion that the effect is small, nor have they done so elsewhere as of this writing. W2018 also does not supply evidence from any benchmark study run in 2011, when the bug was discovered, to support the comparison of its magnitude to the effects of light-curve sampling error, a real physical effect discussed in section 3.5. The latter error, in any case, does not negate the effects of the bug on diameter results, which the curve-fit analysis presented in M2018b suggests is significant. The claim quoted above cannot be evaluated until NEOWISE identifies the benchmark asteroids and detailed study results that were used to quantify the effects of the bug.

W2018 attempts to characterize the importance of the bug in its Figure 3:

> As shown in our Figure 3, these two objects are extreme examples of the software issue; it has a smaller effect on most objects. As described below, the change to the respective diameters of both asteroids is 6% and 8%, which is lower than the minimum 10% uncertainty we claim for our data for the ensemble of objects.

The histogram in W2018 Figure 3 illustrates the effect of the bug on the W3 magnitudes of a small sample of 2170 asteroids, which includes roughly 10% of those objects that were observed in the W3 band. W2018 does not describe whether the 10% subset was selected in an unbiased fashion nor why they did not analyze all such objects or make the underlying data available for others to analyze.

*Response to Wright et al. 2018: Even More Serious Problems with NEOWISE*        Nathan P. MyhrvoldW2018 also offers no justification for why the W3 band alone should be used to evaluate the severity of the bug on NEOWISE results. NEOWISE results to date have *not* included tables of W3 modeled flux; instead it is diameters—and properties like albedo derived from diameters— that are the main result. A figure such as their Figure 3 that purports to address the effect of the bug on *diameters* should show errors in the estimated *diameters*, as well as the effect on magnitudes.

Figure 6 of W2018 does plot diameter errors. The histogram of $(D_{WISE2011} - D_{WISE2018})/D_{WISE2011}$, where $D_{WISE2011}$ is the bug-affected diameter from 2011 and $D_{WISE2018}$ is the corrected diameter obtained from software run in 2018, clearly shows multiple cases in which changes attributable to the bug exceed ±20% (lower-right panel (f) of W2018 Figure 6). The legend in that plot reports that $1\sigma = \pm 6.01\%$. This directly contradicts the claim that the 6% to 8% diameter errors reported for asteroids 25916 and 90367 represent "extreme examples of the software issue." To the contrary, asteroids 25916 and 90367 appear to represent *typical* cases. The bug had a much larger impact than 8% on some diameter estimates.

Of the ~1700 asteroids analyzed to produce Figure 6 of W2018, at least 87 belong to the set of asteroids that M2018b showed had NEOWISE diameter estimates copied from prior ROS sources (section 2.2). It is unclear how Wright et al. treat those 87 cases. Are $D_{WISE2011}$ and $D_{WISE2018}$ both copied from ROS papers? Are both derived using the NEOWISE thermal model? Or is $D_{WISE2011}$ the previously published value (copied from ROS) whereas $D_{WISE2018}$ is a modeled value?

Also unclear is whether, in those cases with multiple estimates from NEOWISE or IRAS, those estimates were averaged prior to comparison, as advocated by W2018. Pre-comparison averaging would tend to decrease the scatter (section 3.3). In either case, the use of simple ratios ignores the fact that each diameter estimate has associated uncertainty (section 3.2). In this case, the ratio distribution is complicated by the possibility of significant correlation between the uncertainties in the pre- and post-bug NEOWISE estimates. A fair and more careful error analysis that accounts for the factors above could find the true impact of the bug to be even more significant than the $1\sigma$ value of ±6% reported in Figure 6(f).

W2018 reports the magnitude by which the bug changed the diameters of asteroids 25916 and 903657, respectively, as "6% and 8%, which is lower than the minimum 10% uncertainty we claim for our data for the ensemble of objects." However, a correct error analysis must *add* the contribution of the software bug to all other sources of uncertainty when computing the total error in diameter. This previously unreported source of error represents a large portion of the claimed overall error budget of ±10% for NEOWISE.

The basis for NEOWISE accuracy claims is a problematic comparison in the ApJ 736 paper of modeled flux that used 1 sample per object per band for 50 ROS asteroids, with observations that were overwhelmingly saturated (sections 2 and 7). W2018 offers no assurance that those modeled fluxes were not affected by the bug.

W2018 and Wright (2019) do highlight incomplete coverage of the light curve as a potentially large source of error, an issue discussed in detail in M2018b and here in section 3.5. Light-curve sampling arising from non-spherical asteroid shapes and asteroid rotation is especially problematic for NEOWISE due to its practice of dividing observations into arbitrary epochs (M2018b). Although W2018 claims that the team was fully occupied in 2011 with quantification of the effects of light-curve sampling, none of the team's publications to date have reported results that quantify the uncertainty due to incomplete light-curve sampling. Wright et al. claim that those effects are much greater than the $\pm 6\%$ effects of the bug, which seems inconsistent with repeated NEOWISE claims of $\pm 10\%$ accuracy if the two effects are additive.

In addressing the finding of M2018b that thousands of NEOWISE model fits miss entire bands of data, W2018 examines two cases and confirms that the poor fits were due to the software bug. Left unanswered, however, is the important question of whether that bug is responsible for some or all of the remaining misses. W2018 does not dispute the results shown in Figures 4 and S8 of M2018b that document poor fits for nearly half the NEOWISE cases.

Whether those poor fits resulted from this bug or some other cause, it is surprising that the team failed to notice until after publication that almost half of the 4-band fits miss an entire band of data. Such a significant oversight calls into question the methods used to assess the quality of thermal-model fits published in the 2011 papers and the scientific management of the mission.

## 4.2 Criticisms of curve fits

W2018 criticizes curve fits presented in M2018b for asteroids 25916 and 90367, which result in values of near-IR albedo $p_{IR}$ that Wright et al. claim is unphysically low: $p_{IR} = 0.012$ for 25916 and $p_{IR} = 0.004$ for 90367. However, it should be noted that NEOWISE results in the PDS archive also include 60 cases where $p_{IR} \leq 0.012$ and 16 cases where $p_{IR} \leq 0.004$. Very low values of visible-band albedo $p_v$ can be found in the NEOWISE result set as well: 61 cases with $p_v \leq 0.012$ and 9 cases with $p_v \leq 0.004$.

The W2018 criticism of these fits misses the point that M2018b was clearly making in its Figure 3, titled "Example models that show poor fit to data." The NEOWISE fit for object 25916 was plotted there to illustrate that it misses *all of the observational data* (as W2018 confirms in its Figure 4a) and is obviously far from a best fit to the data. The example illustrates a serious problem common to a large fraction NEOWISE fits, and it demonstrates how fits that do match the data can be found when using parameters similar to those published for other NEOWISE fits.

It is entirely possible that further investigation or a different analysis of asteroids 25916 and 90367 may well find more realistic values of $p_{IR}$ for these objects and for the other unrealistic values of $p_{IR}$ and $p_v$ found by NEOWISE. However, the matter at hand is not a detailed analysis of these two asteroids—it is the fact that a huge fraction of NEOWISE model fits miss the data they purport to fit.

W2018 claims that they cannot replicate the M2018b fit shown there for asteroid 25916. Ironically, that claim is itself nonreplicable because NEOWISE has never disclosed the modeling details needed to reproduce its fits, nor do they explain how they attempted to replicate the examples from M2018b. As one example of unexplained inconsistency, Figure 4 in W2018 uses different values of *H* for its cases. Elsewhere, W2018 appears to assert without explanation that one can take $p_{IR} = p_v$. Because the NEOWISE model evidently uses so many non-standard methods, it is unsurprising that it fails to replicate results obtained using standard fitting methods. It is also worth noting that M2018b clearly states that it did not use the linear correction to W3 (section 2.3) because that correction had yet to be disclosed by the NEOWISE group.

*4.3 The W2018 conclusion regarding the software bug*

W2018 ends its section 2.2 regarding the software bug and accuracy with this passage:

> Thus we have shown that in spite of a minor software issue that affected early publications, the published WISE diameters are good to within the quoted minimum systematic uncertainty in effective spherical diameter of ~10%.

These points are refuted above in sections 4.1 and 4.2 of the current work, with the accuracy issue also treated further in section 7 below.

As discussed above, Figure 6(f) does not list the bias introduced by the software bug (i.e., the mean of the best fit normal distribution), but rather shows that distribution has $\sigma = 6.01\%$ across a sample of about 1700 asteroids. As discussed in section 3 and sections 4.1 and 4.2, there is every reason to believe that the actual impact of the bug is quite substantial when correctly calculated—i.e., without copied ROS diameters, without averaging multiple estimates NEOWISE prior to comparison, and with ratio distributions that account for the NEOWISE and ROS uncertainty. As a result, the NEOWISE results may need major revision.

# 5. *WISE* estimates of flux uncertainty

*5.1 The double-detection method*

The *WISE* observation pipeline produces images as well as identified sources with estimated magnitude in up to 4 bands, plus estimates of the uncertainty ($\sigma$) in those magnitudes. Hanuš and coworkers (Hanuš et al., 2015) realized that about 10% of each *WISE* frame overlaps with adjoining frames to ensure complete coverage, and this means that objects located within the overlapping portion of the frames are imaged twice in successive frames, with about 11 s elapsing between frames. They realized that this provides a natural experiment for determining whether the uncertainty $\sigma$ predicted by the *WISE* pipeline is correct. Tabulating the Z-statistic for about 400 double detections, Hanuš et al. found that the $\sigma$ were systematically underestimated. The actual value was 1.4 times larger in the W3 band and 1.3 times larger in W4. To date, three NEOWISE group papers have referenced Hanuš et al. (2015), without commenting on this finding (Koren et al., 2015; Mainzer et al., 2015; Nugent et al., 2016).

M2018b extended this analysis to all *WISE* bands and to all of the double detections in the fully cryogenic portion of the *WISE* mission: 17,528 cases in W1; 24,801 in W2; 133,216 in W3; and 102,122 in W4 (see sections 5 and 12.2 in M2018b). With better resolution, M2018b confirmed the finding of Hanuš et al. that the *WISE* pipeline $\sigma$ were underestimated, reporting actual values that were 2.49 times larger in W1, 1.47 times larger in W2, 1.565 times larger in W3, and 1.27 times larger in W4 (see Figure 2 in M2018b). M2018b also found that the Z-statistic for the pairs of detections was non-Gaussian and much better fit by a Student's t-distribution. The effect existed for all observations—it was not a result of high-noise measurements—and the effect *increased* when double detections were filtered to include only high-SNR observations. M2018b also verified that this phenomenon is not unique to asteroids and occurs as well for millions of double detections of stars observed during the fully cryogenic mission (see Figures S4 and S5 in the Supplemental Information of M2018b).

The findings of Hanuš et al., and the extension of that work in M2018b, are of great importance to any error analysis of the *WISE* data. The NEOWISE results papers performed error analysis by using a Monte Carlo simulation that assumed that all observational errors conformed to a Gaussian distribution having a standard deviation based on the *WISE* pipeline $\sigma$ (Mainzer et al. 2011c, Masiero et al. 2011). Because the input to the error analysis includes estimated errors that were grossly underestimated by the *WISE* pipeline, the output of the error analysis must also underestimate error in diameter and other parameter estimates (see section 5 in M2018b).

W2018 now attempts to refute this work in its section 3.1, though it has little to say about the central issue. Instead it focuses considerable rebuttal attention on whether the distribution of errors is perfectly Gaussian or not. That question is tangential to the issue raised in M2018b for the NEOWISE error analysis, which is that it assumed *both* the wrong distribution *and* the wrong variance. As discussed in M2018b, it is the combination that is critical, not the distribution issue alone.

W2018 misstates the theoretical basis for Monte Carlo simulations when it asserts that the NEOWISE Monte Carlo error analysis is insensitive to the distribution of errors. A Monte Carlo simulation for error analysis draws its input from distributions that represent the system's inputs and their measurement errors. An accurate simulation requires sampling distributions for the inputs that have the same variance as one expects in the system being simulated. Ideally, one also uses the most closely matching distribution. The normal or Gaussian distribution is often used as a proxy for the correct distribution, but it is well known to provide a poor approximation if the true distribution has heavier tails than a Gaussian (Feigelson and Babu, 2012; Ivezić et al., 2014; Wall and Jenkins, 2012).

If the error analysis is based on a standard deviation $\sigma$ that is too small by a factor of 2.5 (as it appears to have been with band W1), then the analysis will almost invariably give the wrong answer. Using a much fatter-tailed Student's t-distribution with a higher variance, instead of a Gaussian with lower variance, can similarly generate erroneous results. W2018 claims, in effect, that the actual size of the *WISE* observational error does not matter to the error analysis. That cannot be true for a valid error analysis (Dieck, 2017; Taylor, 1997).

W2018 finds no flaws with the approach of looking at the variance in repeated observations of the same object. Indeed, Wright et al. (2010) used this same standard method:

> The individual frames are analyzed to provide astrometric and photometric information. Analyzing the scatter among the values from individual frames gives the noises shown in Figure 9. The dashed line shows a 5:1 SNR in 8 frames scaled to the 11-frame case plotted, giving σ(m) = 0.185 mag where n is a noise term fitted to the data points.

That paper includes a figure presenting the results with this caption:

> Fig 9—Standard deviation of *WISE* magnitudes derived from the repeatability on sources seen in 11 frames. The standard deviations have been binned into bins with width 0.2 mag, and the resulting histograms have been fit with a gaussian using outlier rejection.

Repeated observations of sources across 11 frames generated a distribution that was stated to be approximately Gaussian. This differs from the approach of Hanuš et al. (2015) and M2018b principally in looking at variance across pairs of frames, not 11 frames as in Wright et al. (2010), and in using "outlier rejection." Indeed, it is possible that the Wright et al. 2010

"outlier" rejection was overly aggressive and threw out valid data that would have alerted them to the issue.

The primary goal of the *WISE* mission was to produce an atlas of the sky in 4 bands based on 8 to 12 co-added frames at each location (Wright et al. 2010). The additional frames provide a much larger sample that better approximated the underlying distribution and converged to expected values. Asteroids observations, in contrast, must rely on single frames, and the *WISE* pipeline estimated those errors from small samples.

As the Appendix to M2018b explains, the Student's t-distribution arises from small-sample-size estimates of the mean and variance of a Gaussian distribution. The single-frame *WISE* observations are just that: estimates of the underlying pixel statistics made using very low sample sizes from the pixel statistics on a single frame. As with other pixel-counting statistics, those likely follow a Poisson distribution rather than a Gaussian, but they are close enough to mimic the Student's t-distribution.

This theoretical finding of M2018b explains the source of the underestimate—the *WISE* pipeline generated small-sample-size estimates of the uncertainty rather than the true uncertainty. The small sample sizes inevitably create heavier-tailed distributions (in the manner of Student's t-distribution). The approach to calculating the *WISE* pipeline $\sigma$ is of such small sample size that it is misleading to use it as the uncertainty. Conversely, interpreting the *WISE* pipeline $\sigma$ as the standard deviation of a Gaussian distribution, as done by NEOWISE, will seriously underestimate the errors.

W2018 finds no fault with the double detection logic or Z-statistic. Wright et al. instead attempt to use an analysis of the *WISE* $\sigma$ statistics on stars to rebut the findings of Hanuš et al. 2015 and M2018b. However, M2018b already correctly analyzed both double-detection statistics and repeated measurements of ~1 million stars used for SDSS calibration (see sections 5, Appendix, and Supplemental Information in M2018b). W2018 takes no issue with that analysis.

*5.2 Analysis of ecliptic polar-star observations*

Rather than investigating the estimation of uncertainty by the *WISE* pipeline, W2018 offers an analysis that contributes little or nothing to that question.

> To get a very dense sample of data to investigate these claims, we have collected data on stars close to the ecliptic poles that were observed dozens of times during the 4-band cryogenic mission. …
>
> First, we compute the rms of the flux distribution for each star, and then compute the mean of the rms values in each flux bin as a measure of the characteristic dispersion in flux measurements. Second, we compute the mean of the *wNsigflux* values reported by the noise model in the WISE database. These two quantities are compared to confirm the accuracy of the noise model.

Their Figure 8 plots the average (rms) scatter against the mean of the error estimates. Such a method might identify a systematic bias in the error estimates, but it is *not* sensitive to the issue claimed by Hanuš et al. (2015) and M2018b, viz. that there is a systematic underestimation of the uncertainty on an individual basis. In particular, thermal modeling does *not* involve the rms scatter, nor does it use the mean uncertainty. Instead models make use of observation point-by-point fluxes and uncertainties, which is why Hanuš et al. 2015 and M2018 focused on those.

The relevant measure here is the statistic $\hat{f}_i$, which is known in statistics as the Z-statistic, Z-score, or standardized score. In the case of repeated flux measurements $f_i$, each with estimated uncertainty $\sigma_i$, $\hat{f}_i$ is given by

$$\hat{f}_i = \frac{f_i - \bar{f}}{\sigma_i}, \qquad \bar{f} = \frac{1}{N}\sum_{i=1}^{N} f_i. \tag{4}$$

If the errors in the $f_i$ are normally distributed with standard deviation $\sigma_i$, then the $\hat{f}_i$ should have unit standard deviation if the $\sigma_i$ correctly describe the variance in the $f_i$. If the standard deviation is not unity, then the $\sigma_i$ do not accurately reflect the variance of the distribution that the errors are drawn from. This test is not sensitive to the distribution being precisely normal, and it holds under the same conditions as one would expect the central-limit theorem to apply (Bevington and Robinson, 2003; Bonamente, 2013).

W2018 does not identify the stars it uses, but stars and other sources near the ecliptic poles were used by Jarrett et al. (2011) in calibrating the *WISE* zero points, in conjunction with spectra from the *Spitzer Space Telescope* and other sources. As M2018b notes, we should expect different pixel statistics from stellar sources because their temperatures are in general much higher, and they are thus dimmer in the longer-wavelength W3 and W4 bands. As a result, stars are not ideal as asteroid proxies, but they do offer repeated observations at far higher repetition count than the double detections studied in Hanuš et al. (2015) and M2018b. Rather than two observations within 11 s, we get thousands of observations about 90 min apart.

Jarret et al. (2011) provide a list of sources, and Table 3 below lists 19 of those that have the largest observation counts, typically 2200 to 3200 observations in bands W1, W2, and W3, with generally far fewer in W4.

| Name | Observation Counts | | | |
|---|---|---|---|---|
| | W1 | W2 | W3 | W4 |
| NGC6552 | 0 | 2531 | 2750 | 2657 |
| KF03T1 | 3092 | 3148 | 3135 | 395 |
| KF03T2 | 3088 | 3134 | 3141 | 591 |
| KF06T1 | 2431 | 2482 | 2218 | 152 |
| KF06T2 | 3100 | 3162 | 2361 | 225 |
| KF06T3 | 3104 | 3134 | 3109 | 247 |
| KF03T4 | 3087 | 3120 | 3062 | 425 |
| KF05T1 | 3100 | 3134 | 2984 | 36 |
| KF02T1 | 3098 | 3134 | 3152 | 744 |
| Bp66_1073 | 3076 | 3127 | 3127 | 2462 |
| KF02T3 | 3108 | 3142 | 3124 | 328 |
| HD270422 | 2710 | 2724 | 2745 | 2507 |
| HD270467 | 2765 | 2800 | 2807 | 460 |
| 5581475 | 2700 | 2791 | 637 | 125 |
| WOH_G642 | 2766 | 2789 | 2795 | 353 |
| HD41466 | 2714 | 2739 | 2750 | 854 |
| WOH_S527 | 2730 | 2766 | 2803 | 2355 |
| HD270485 | 2708 | 2743 | 2802 | 291 |
| HD271776 | 2729 | 2769 | 2773 | 185 |

**Table 3. Observation counts for ecliptic polar sources from Jarret et al. (2011).** The observations were filtered so that $nb = 1$ to avoid problems with source blending and $w1rchi2 < 10$ (and similarly for W2, W3, and W4) to discard most particle hits and resolved sources. The *WISE* pipeline already performs a filter rejecting observations closer than 50 arcsec to a frame edge. Data filters on $ssa\_sep > 0$ and $dtanneal > 2000$ were checked and rejected no observations. The counts give the number of observations in each band.

W2018 lists several data-quality filters for the stellar sources it uses, namely that $ccflags = 0$, the blending flag must be $nb = 1$, the per-band point spread function $\chi^2$ parameters $w1rchi2, w2rchi2, w3rchi2, w4rchi2$ must satisfy $wxrchi2 < 10$, the South Atlantic Anomaly flag must be $saa\_sep > 0$, and sources within 50 arcsec of the frame edge were not used.

Of these flags and filters, however, prior NEOWISE publications mentioned only the $ccflags$ criterion. Jarrett et al. (2011) made no mention of the other criteria either. Ignoring sources closer than 50 arcsec to the edge of the frame is already built into the *WISE* pipeline (Cutri et al., 2011). If all NEOWISE publications used the combination of filters described in W2018, they neglected their obligation to document that practice. Alternatively, if W2018 now prescribes a different data-quality standard than that typically used by NEOWISE, that too raises ethical questions.

The analysis presented here adopts $ccflags = 0$, $qual\_flags$ = A,B,C, and $wxrchi2 < 10$. The filter $dtanneal > 2000$ s on the time since last detector anneal implies that this parameter is of little importance to observations near the ecliptic pole. Twice a day, *WISE* anneals the W3 and W4 detectors (Cutri et al., 2011), during what would otherwise be a polar observation because that area is oversampled. Observations during the annealing process are not recorded. After annealing, the time for the orbit to reach the opposite pole is about 44 minutes, while the 2000 s limit equates to just 33.3 min. Consistent with this, the polar source observations shown here have $635.2 \leq dtanneal \leq 77875.3$ sec, so this filter removed no data sources. The South Atlantic Anomaly separation variable was in the range $12.5 < saa\_sep < 120.8$, so the filter requiring $saa\_sep > 0$ also did not eliminate any observations. The net result is that the data quality restrictions used here are the *same* as those described in W2018.

The observation counts in Table 1 reflect application of these filters. Figure 7 plots the filtered observations for an example source, the star WOH_S527, which was observed 2355 to 2803 times, depending on the *WISE* band. The top row of Figure 3 presents actual data points with error bars, while the bottom row plots histograms of the Z-statistic from Equation (1).

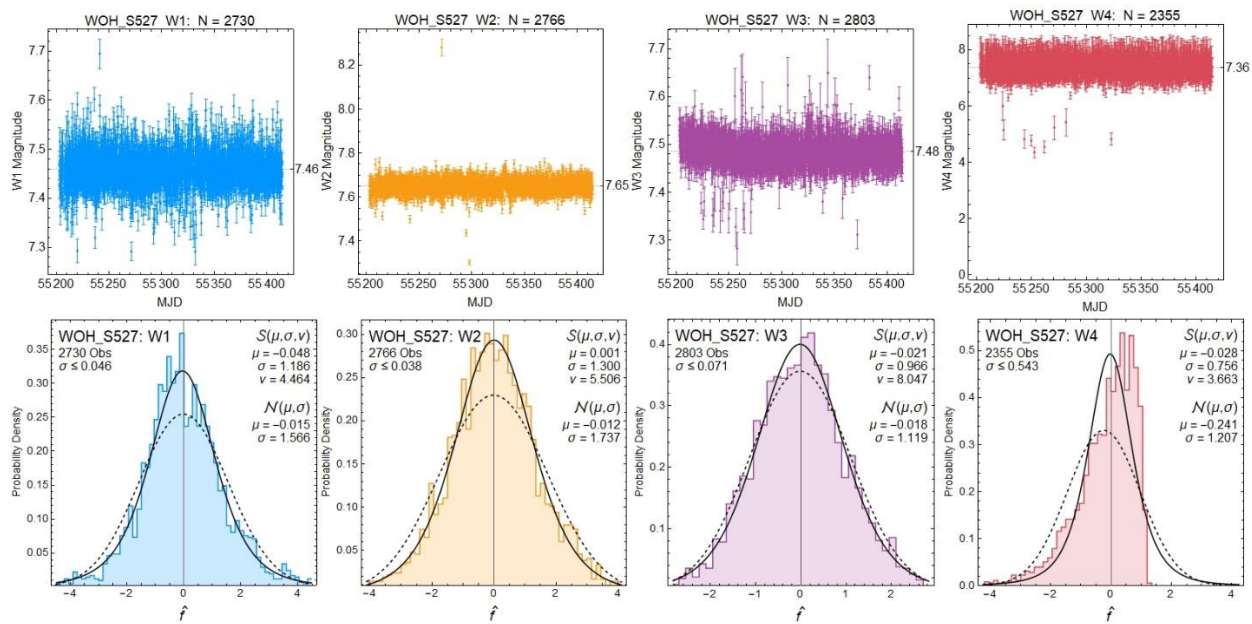

**Figure 7. Example of repeated *WISE* observations of a stellar source.** Top row: repeated observations of star WOH_S527 in all 4 *WISE* bands, with error bars indicating the *WISE* pipeline $\sigma$ for each observation. Horizonal lines and labels mark the means. Bottom row: corresponding histograms of the Z-statistic $\hat{f}$ of Equation (1) and the best-fitting normal distributions $\mathcal{N}(\mu, \sigma)$ (*dashed black line*) and fit parameters. In all 4 bands, $\sigma > 1$ for the best-fitting normal distribution, indicating that the per-observation $\sigma_i$ underestimate the true uncertainty. Also shown are the best-fitting Student's t-distributions $S(\mu, \sigma, \nu)$ (*solid black line*) and associated parameters. All best-fit distributions were found by using maximum likelihood estimation.

As the histograms of Figure 7 illustrate, in all bands the best-fitting normal distribution fits the data very poorly. Standard deviations exceed unity and indicate that the single frame *WISE* pipeline $\sigma$ underestimate the true uncertainty by factors ranging from 1.12 (for W3) to 1.74 (for W2). The per-band factors differ from those found by Hanuš et al. 2015 and M2018b because they depend on pixel-counting statistics that vary with the brightness in band and thus the temperature of the source, as explained in M2018b.

Figure 8 plots histograms of pooled Z-statistics from all 19 sources in Table 3. Here, too, the normal distribution fits poorly, and the standard deviations of the best-fitting normal distributions reveal that the *WISE* pipeline $\sigma$ underestimates the true uncertainty by factors of 1.37 to 2.17. In both Figure 7 (*bottom row*) and Figure 8 (*top row*) the Student's t-distribution obviously offers much better fits to the distributions of the Z-statistic than normal distributions do. Figure 9 explores this further with P-P probability plots, which plot the cumulative distribution of the best-fitting theoretical distribution versus the cumulative distribution of empirical data.

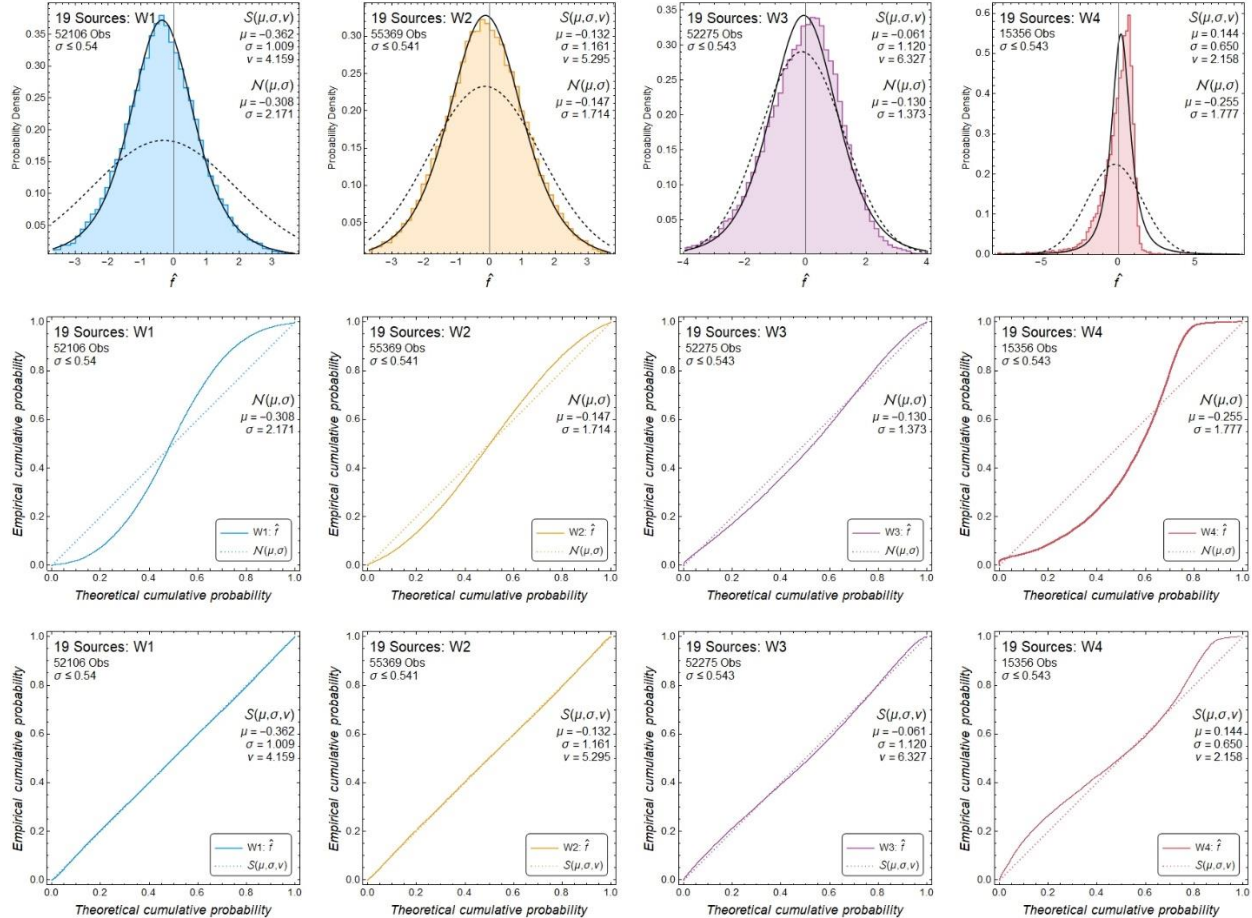

**Figure 8. Analysis of the Z-statistic for repeated sources.** Top row: histograms of the Z-statistic of Equation (1) for observations pooled from the sources of Table 1. The best-fit normal distributions ($\mathcal{N}(\mu,\sigma)$) (*dashed black line*) and Student's t-distributions ($S(\mu,\sigma,\nu)$) (*solid black line*), as determined by maximum likelihood estimation, are shown along with corresponding fit parameters. In all bands, the $\sigma > 1$ for the best-fitting normal distributions, indicating that the per-observation $\sigma_i$ underestimate the true uncertainty. Middle row: P-P probability plots for the best-fit normal distributions (*dotted line*) along with the actual Z-statistic $\hat{f}$ (*solid line*). Bottom row: P-P probability plots for Student's t-distribution. The P-P plots reveal that normal distributions are poor fits for all bands, with the best fit occurring in band W3. The Student's t-distribution, in contrast, provides is excellent fits to bands W1, W2, and W3, but a lower-quality fit for band W4, which has very few data points compared to the other bands.

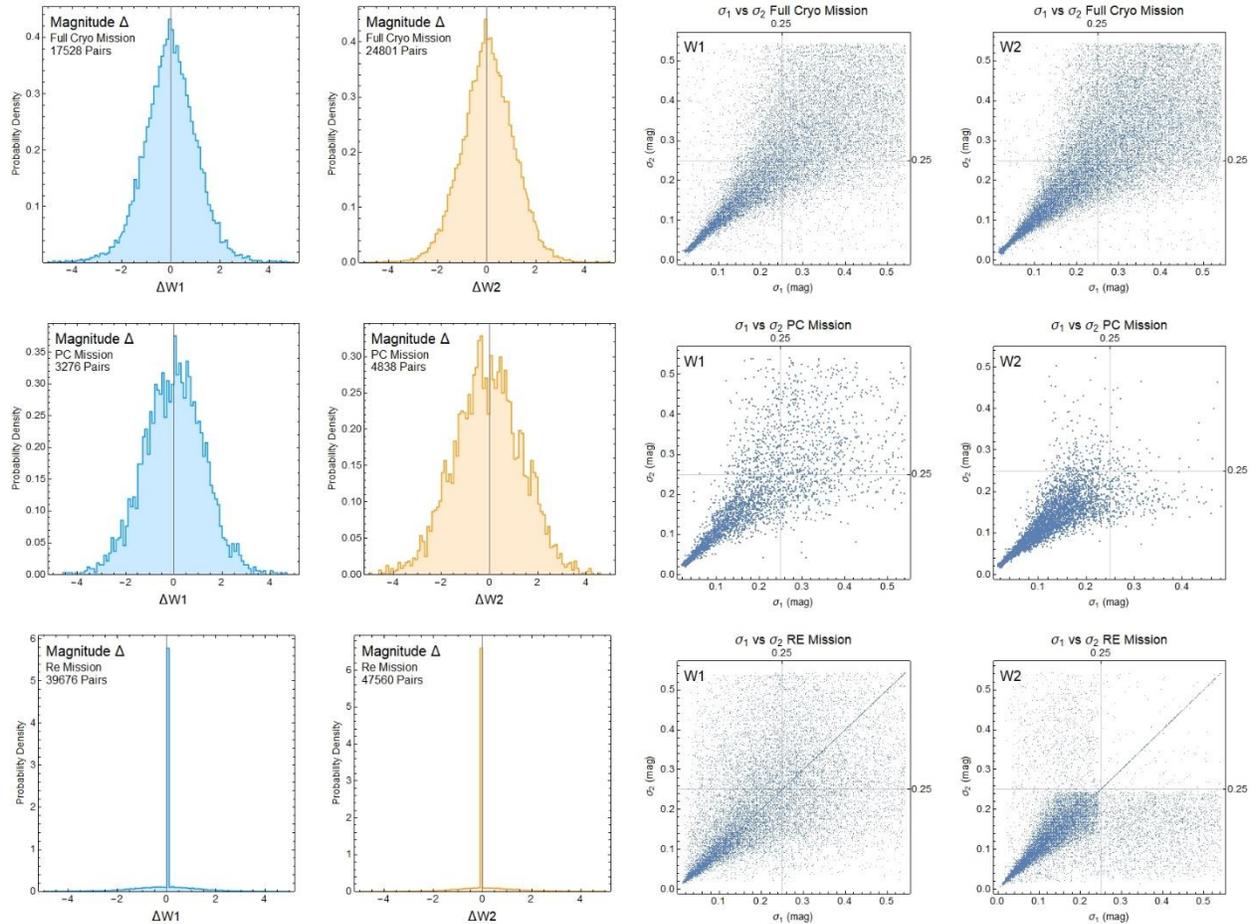

**Figure 9. Change in *WISE* pipeline processing during the reactivation mission phase.** *WISE* images some asteroids twice separated by ~11 s. These "double detections" offer a way to test the *WISE* pipeline uncertainty estimates $\sigma_1, \sigma_2$. Histograms of the differences in *WISE* magnitude found in double detections in bands W1 and W2 for the fully cryogenic (Full Cryo) portion of the *WISE* mission are shown (*top row, left of center*) along with scatter plots of the differences in $\sigma_1, \sigma_2$ for each of the double detections (*top row, right of center*). The same statistics are plotted for the post-cryogenic (PC) mission (*middle row*). While the PC mission plots are qualitatively similar to the full-cryo plots, those for the reactivation (Re) mission (*bottom row*) clearly illustrate a shift, as was reported in M2018b. During this most recent mission phase, fluxes were set equal in the vast majority of double-detection cases. The scatter in $\sigma_1, \sigma_2$ was also modified (*bottom row, right of center*): $\sigma_1 = \sigma_2$ for many more cases than in prior mission phases. An irregularity is apparent in the quadrant where $\sigma_1 > 0.25$ and $\sigma_2 > 0.25$.

The P-P plot shows that the Student's t-distribution provides an excellent fit to bands W1, W2, and W3, but a poor fit to W4. The normal distribution, in contrast, is inferior in every band, but comes the closest to matching the fit in W3. The P-P plots demonstrate that this phenomenon is not the result of isolated outliers because the empirical distribution of the Z-statistic is well fit by the Student's t-distribution across the entire range of the distribution. These results confirm the findings of M2018b and refute W2018.

The W2018 analysis relating rms scatter to mean uncertainty $\sigma$ is not relevant to the findings here. At best, the rms scatter and mean uncertainty are related to the location parameter of the Z-statistic empirical distribution—i.e., the $\mu$ parameter of the normal or Student's t-distributions, rather than the shape parameters $\sigma$ and $\nu$. That parameter captures the average bias in the estimates of uncertainty. While the results here show that bias to be nonzero, it is small and not the primary focus of the analysis. Note also that for the normal distribution, the $\sigma$ parameter is the standard deviation, but that is not true of the Student's t-distribution.

The fact that neither the normal distribution nor a Student's t-distribution provide good fits to the W4 band is likely related to relatively poor W4 statistics due to the generally high effective temperature of stars, and hence very little flux in the W4 band. Proper study of this effect is beyond the scope of this article. The crucial point at present is that the W4 distribution nevertheless demonstrates that the *WISE* pipeline $\sigma$ underestimate the true uncertainty.

The original insight of Hanuš et al. (2015) that double detections provide a natural experiment for assessing the accuracy of the *WISE* pipeline $\sigma$ is valid. Figures 7 and 8, along with Tables 3, A1, and A2, demonstrate that repeated observation of the *WISE* calibration stars from Jarret et al. (2011) reveals the same phenomenon documented by Hanuš et al. (2015) with asteroid double detections: the *WISE* pipeline $\sigma$ values are too small, and the distribution has much fatter tail, like that of a Student's t-distribution.

## 5.3 Changes in the reactivation mission

In the post-cryogenic (PC) and reactivation (Re) portions of the *WISE* mission, observations were collected only in bands W1 and W2. M2018b observed that during the Re phase, the *WISE* pipeline seemed to treat double detections of asteroids differently than it had before. W2018 disputes this, on the basis of a "cursory check" of 12 cases for asteroids and with a "random selected scan" (their Figure 9) from late 2014 (during the Re mission) of stars, not of asteroids.

Figure 9 plots histograms and scatter plots of all double detections of *asteroids* in the W1 and W2 bands during the PC and Re mission phases. The figure also includes a comparison to corresponding plots for the fully cryogenic phase for the same bands. The PC plots are qualitatively similar to the full-cryo plots.

However, it is obvious that a profound change occurred for double detections in the Re mission phase. In particular, the case where both detections have the same value ($\Delta = 0$) became vastly more common. A qualitative change also clearly occurred in the *WISE* pipeline estimates of uncertainty $\sigma$. Although cases remain where the *WISE* magnitudes or $\sigma$ differ between the two observations in a double-detection pair, the statistics reveal a dramatic qualitative shift.

The "cursory" examination of 12 asteroid double detections reported by W2018 was clearly insufficient to reveal the phenomenon; while the Re mission phase did gather some double detections having different flux and different estimated uncertainties, those were vastly outnumbered by the altered cases that were clearly set equal (Figure 9, bottom row).

Rather than check a large sample of asteroid double detections, the analysis performed by W2018 in its Figure 9 was of double detections of *stars*. It is thus irrelevant to the topic of M2018b's finding about changes in the NEOWISE asteroid pipeline.

If, however, nothing changed in the Re mission phase for stellar observations (as asserted in W2018), the dramatic shift for asteroids demands explanation. A change in the underlying instrument, or in lower-level parts of the *WISE* pipeline, would presumably affect all sources.

*5.4 The W2018 conclusion regarding WISE flux errors*

W2018 ends its section 3.1 with this passage:

> Thus, we have refuted the claim of M2018b that the WISE measurement uncertainties are improperly estimated and used, and that the WISE pipeline was changed for the NEOWISE reactivation mission.

These points are thoroughly refuted above in sections 5.1 through 5.3. M2018b demonstrates the double-detection analysis for all SNR, and then takes subsets. It shows that the problem exists at *every* SNR level for double detections of asteroids and stars. The theoretical explanation provided by M2018b also applies to every SNR level.

Section 5.2 establishes that the same effect occurs for repeated *WISE* observations of ecliptic polar sources. There is no SNR-specific effect as far as M2018b or the current work are concerned.

The notion that no NEOWISE diameter estimates were affected by the misestimated flux uncertainty is also unsupported. NEOWISE papers, as documented in the appendix and supplemental online information to M2018b, make a variety of ad hoc changes to the flux uncertainty. However, it is far from clear that this eliminates all impact on the diameters. Until that is demonstrated by reproducible calculations, the claim of no diameter impact must be treated as unproven. In contrast, it is clear that if the flux uncertainties were underestimated, then the Monte Carlo estimates of diameter uncertainty were underestimated as well.

Estimation of observational uncertainties is a fundamental part of the observing task, and the misestimated flux uncertainties have been a barrier to use of the WISE/NEOWISE dataset outside the NEOWISE group. Indeed, the double-detection analysis of Hanuš et al. 2015 was conducted as part of a thermophysical modeling effort that used the WISE/NEOWISE data and ran into difficulties with the flux uncertainties as published.

# 6. The relationship between $H$, $p_v$, and $D$

*6.1 Understanding NEATM and the definition of $p_v$*

Equation 1 in many of the NEOWISE papers is the classic relationship between geometric albedo in the visible band $p_v$, asteroid diameter $D$, and visible-band absolute magnitude $H$ for a spherical asteroid. It is also equation 1 in M2018b, where it is expressed as the definition

$$p_v \equiv \left(\frac{1329}{D}\right)^2 10^{-\frac{2H}{5}} \quad . \tag{5}$$

M2018b reported that, although many of the NEOWISE papers invoke Equation (5), about 15,000 of the NEOWISE results violate this relationship, sometimes by large margins. In some of the most extreme cases, $D$ is off by a factor of 4; discrepancies in $p_v$ range up to a factor of 24 (see Section 7 and Figure 6 in M2018b).

The W2018 section 3.3 "Should $H$ be fit exactly?" attempts to address this issue. Because $H$ is a measured quantity and NEOWISE does not operate in the visible, the NEOWISE studies rely on *H* values from the Minor Planet Center (MPC). NEOWISE thus shouldn't be fitting *H* at all. W2018 comments about Equation (5) (M2018b Equation 1) that:

> One should note that this is *not* the definition of *H*. *H* is the visual magnitude of the object observed at zero phase angle from a distance of 1 AU, when it is 1 AU from the Sun (Bowell et al. 1989).

Of course, Equation (5) is a definition of $p_v$ as a function of *H* and *D*—that is what "$p_v \equiv \cdots$" means. Nowhere does M2018b assert or imply that Equation (5) is a definition of *H*. In fact, as the notation unambiguously states, it *is* the definition of geometric albedo for a spherical asteroid.

The albedo $p_v$ appears in the NEATM as a component of the sub-solar temperature $T_{ss}$ (Harris, 1998), which is a function of the distance $r_{as}$ of the asteroid from the sun, in AU. $T_{ss}$ also depends on parameters such as the phase integral $q$, the NEATM emissivity $\epsilon$, and the beaming parameter $\eta$, as well as the solar constant $S$, and the Stefan-Boltzmann constant $\sigma$.

$$T_{ss}(r_{as}) = \left(\frac{S(1 - p_v q)}{\epsilon\, \sigma\, \eta\, r_{as}^2}\right)^{0.25} = \frac{T_1}{\sqrt{r_{as}}} \qquad (6)$$

As Myhrvold (2018a) explores in depth, one can reparameterize the NEATM $T_{ss}$ function in terms of the new parameter $T_1$, as shown in Equation (6). A physical interpretation of $T_1$ is the NEATM sub-solar temperature at $r_{as} = 1$. This parameter replaces both $p_v$ and $\eta$, which appear nowhere else in the NEATM.

This new parameterization has both conceptual and numerical advantages. Neither the albedo $p_v$ nor the absolute magnitude *H* are required to perform thermal modeling or to obtain a diameter estimate from the NEATM. Indeed, *H* appears in NEATM only insofar as it is a component of $p_v$, and $p_v$ appears only in Equation (6) in combination with the beaming parameter $\eta$. *H* is required, however, when one wishes to obtain $p_v$, as it appears in the definition of $p_v$, Equation (5).

While the new parameterization makes this explicit, it is simply making clear a mathematical property of NEATM *that was always present*. Thus, regardless of whether one uses the new parameterization, the mathematical structure of Equation (6) demonstrates that in NEATM, *H has never been required for obtaining a diameter estimate*.

This is proven mathematically in Myhrvold (2018a), using simple algebra; but the essence is clear. We know that, by definition, the Bond albedo $A_v = p_v q$ must obey the relationship $0 \leq A_v \leq 1$. If this were not the case, then $T_{ss}$ could become complex. Similarly, we know that the emissivity term in Equation (6) obeys $0 \leq \epsilon \leq 1$. Meanwhile, the beaming parameter $\eta$ is not limited, so $0 \leq \eta \leq \infty$. Thus any possible value of $p_v$ can be compensated by $\eta$. This is why it is possible to replace $T_{ss}$ with the reparameterization, but it also guarantees that in a correct implementation, *H* cannot affect *D*.

Harris and Harris (1997) derived a correction formula for changing estimates of $p_v$ due to improvement measurements of *H*. They presented their approach as an approximation based on the assumption that $(1 - A_v)D^2$ is approximately independent of *H*. That insight is broadly true for thermal models, but the context here is that Harris and Harris (1997) pre-dated the introduction of the NEATM (Harris, 1998). The reparameterization of Myhrvold (2018a) shows that in the specific context of the NEATM, the adjustment of Harris and Harris is exact.

Like any observed quantity, *H* is subject to observational errors, but it also follows from Equation (6) that errors in *H* only affect the value of $\eta$ only—they *cannot* affect the estimate of *D*, so long as we believe that $0 \leq A_v \leq 1$. There is therefore no reason that NEATM estimates of *H*, *D*, and $p_v$ should ever violate Equation (5). Obviously, errors in *H* can have a direct impact on $p_v$ via Equation (5). But for any given pair *H*, *D* then Equation (5) should hold for $p_v$.

Prior to the introduction of this method in Myhrvold (2018a), NEATM modelers typically used the full expression for $T_{ss}$, with $p_v$ being expressed in terms of *H* and *D*, per Equation (5) (Delbo et al., 2015; Delbó et al., 2007; Harris, 2005, 1998; Harris and Drube, 2016; Harris and Lagerros, 2002; Kim et al., 2003; Rozitis and Green, 2011; Trilling et al., 2008). The parameters to be fit by the model-fitting algorithm are then *D* and $\eta$. With care to avoid numerical issues with the fitting algorithm, this approach can produce good results. Having obtained *D* in this manner, *H* would be used to calculate $p_v$ via Equation (5), so that relationship should never be violated.

## 6.2 The approach of W2018 in fitting H

The heart of the dispute about Equation (5) is that W2018 asserts that the visible-band albedo *H* must be fit as part of thermal modeling. W2018 presents a model with a statement that:

> The thermal fit model gives the predicted value for the absolute magnitude, $H_p$ and the term $\left((H - H_p)/\sigma_H\right)^2$ contributes to the $\chi^2$ when fitting a model.

This statement demonstrates extreme confusion about the basic mathematics and physics behind the NEATM, and also about the definition of $p_v$ in Equation (5). The statement seems to confuse the model's inputs with its outputs.

The NEATM is a specific model (Harris, 1998) that NEOWISE papers state that they use. In the NEATM, the value of *H* is an observed *input* parameter that affects the value of $\eta$. *H* is not an output of the model and the NEATM, as conventionally known in the thermal-modeling community, *does not predict H*!

Physics intuition can also help clarify this crucial point. To "predict" *H* is not possible when using a model like the NEATM (Harris, 1998) that deals exclusively with emitted *thermal* radiation. There is no mechanism within the NEATM to predict how bright an asteroid is in the visible (i.e., its *H*) simply by using *WISE* infrared-band observations as the input.

One could of course imagine physical models that *could* predict *H*, perhaps by making assumptions about a constraint on $p_v$ provided by $p_{IR}$. However, such a model could not be called the NEATM. Neither W2018 nor the original NEOWISE papers explain any such rationale or method.

The clear answer to the title of section 3.3 of W2018 is that NEOWISE should not be fitting *H at all*. If the team did "fit" *H* as part of its analytical process, it did so in violation of the nature of their instrument and numerous direct statements that they were using the NEATM model.

The NEOWISE papers stated that they obtained the *H* values for asteroids from the literature or from the Minor Planet Center (Grav et al., 2011b; Mainzer et al., 2011b, 2011a; Masiero et al., 2014, 2011). They also stated that *H* and its uncertainty were used in their Monte Carlo error analysis. They did not disclose in any paper that *H* was an *output* from the model. W2018 now states that it was.

As shown in Myhrvold (2018a) and mentioned above, *D* does not depend on *H* in a correct NEATM implementation. In an incorrect implementation, however, anything is possible. Despite making the assertion that *H* should be fit, W2018 nevertheless does not disclose the actual models used in NEOWISE papers to fit *H*. Instead, they present a highly simplified model of a circular disk to illustrate presumably similar principles.

The Monte Carlo calculation that W2018 offers in its Equation (4) is an incompletely described model of a flat disk. Myhrvold (2018a) offers such a model in its section 2, and Equation (3) of that paper appears to be equivalent to W2018 Equation (4), with appropriate adjustments in notation (i.e., using $\epsilon(\lambda)$ rather than $1 - p(\lambda)$).

W2018 intends its model to be based on 5 observational bands; the *WISE* bands W1 through W4, plus a fifth visual band V. Wright et al. also assume that the albedo is identical in all bands, i.e., $p(\lambda) \equiv p$ for a constant $p$. W2018 goes on to claim that this shows that *H* must be fit, on the basis that the total $\chi^2$ value will involve *H*.

Their assumption that independent V band observations are taken at the same time as the *WISE* observations does not hold true for *WISE*. If *WISE* had a visible-band sensor, then fitting *H* based on that sensor would be legitimately be part of the *WISE* data analysis, and a fitted value of *H* in that case would legitimately be an output of the analysis.

*WISE* did not have a visible-band sensor, however. The overwhelming conventions of scientific data analysis are that one should in that case find *H* from the literature and treat it as an *input* to the model (with its estimated uncertainty), not an *output* of the model. This is also what the NEOWISE team stated they did in their papers.

The standard practice reflects the reality that the astronomers in the best position to understand the various effects that determine *H* are those who observed it directly. In particular, the phase angle of the observations are critical, as is the slope parameter *G* in the HG phase law (Buchheim, 2010). Because *WISE* lacks visible-band sensor, it does not record the phase angle at the time of observation.

If NEOWISE intended to make a full meta-analysis of the *H* values from visible-band observations, as has been done in other studies (Veres et al., 2012; Vereš et al., 2014), then it would need an extensive amount of additional data and analysis, plus documentation of that procedure in its publications.

Thus, while W2018 claims that *H* must be fit because the total $\chi^2$ value involves *H*, that is the case only because their model was unrealistically constructed precisely so that this must happen.

Having used visible-band observations V to set up the problem, W2018 quickly switches from observed fluxes from the sun $F_\nu^\odot$, restricted to the visible band, to the parameter *H*. The switch cannot mask the invalidity of their approach. If one makes multiple visible-band flux measurements V from a sensor, then *H* is legitimately an output—presumably based on the mean value of V, with appropriate adjustments for the phase model and for distances from asteroid to observer, and asteroid to Sun (Buchheim, 2010). But a hypothetical instrument measuring visible-band flux has little to do with fitting $H$ with as single "observation" obtained from the literature.

A second unrealistic feature of this model in W2018 is its assumption that the albedo $p$ is the same in every band. This premise—that $p_v = p_{IR} = p$—links the hypothetical V-band estimate of $D$ to the IR-based estimates of $D$ from W1 through W4. If one uses a set of W1...W4, V synthetic fluxes having random errors to fit the model, then it is true that one could "fit" $H$. However, that outcome results from contrived and unrealistic assumptions. It does not justify using *WISE* fluxes to fit $H$.

The treatment of albedo in W2018 echoes a similar phenomenon across the NEOWISE studies. The earliest studies, including Masiero et al. 2011, fit the near IR albedo $p_{IR}$ and assumed it to be the same in both the W1 and W2 bands. Later, Masiero et al. 2014 fit separate IR albedos $p_{IR1}, p_{IR2}$ in W1, W2 for a subset of the asteroids. Because the extra degrees of freedom allowed $p_{IR1}, p_{IR2}$ to vary independently, they found different values for those albedos than had been found for the single albedo $p_{IR}$.

Indeed, this is exactly what one would expect on general grounds—adding extra degrees of freedom generally makes it easier to fit a model. This does not, by itself, mean that it is justified to add the extra degree of freedom. In the case of IR albedos $p_{IR1}, p_{IR2}$, there is a physical reason to expect that they might be different. But the appropriate statistical technique in such a situation is to use a model comparison criterion, such as $AIC_c$ or BIC (Feigelson and Babu, 2013; Ivezić et al., 2014), to ensure that the extra degree of freedom contributes enough additional information to be statistically justifiable.

In the example of W2018, they attempt to justify adding a degree of freedom for $H$ on the basis that doing so gives them results of lower $\chi^2$ and, as a consequence, less scatter in the fit results for $H$. This is clearly circular reasoning.

In a model with statistical noise—i.e., the observational uncertainty in the fluxes—one would always expect that extra degrees of freedom would be used by the system, in particular to obtain a better fit. Equation (4) of W2018 makes an *exact* prediction, when the measurements we apply the model to are *noisy*. Adjusting an additional degree of freedom will essentially always allow us to reduce $\chi^2$. That by itself is *not* justification for adding the degree of freedom.

Indeed, the W2018 approach could be used to justify adding unnecessary degrees of freedom to *any* parameter involved in the equation; one could even "fit" the value of a constant such as $\pi$. In Equation (4) of W2018, $\pi$ occurs as the same value in every band. When one processes synthetic observational data with observational noise added, some of the noise will propagate into the degree of freedom created in the model for $\pi$. When the synthetic fluxes are too low, the model will adjust $\pi$ to be higher, and so forth.

The argument used to justify fitting $H$ in W2018 is simply that the degree of freedom for $H$ is used in the model fitting, an argument that could be used to "prove" that each object in a synthetic set of asteroids ought to have its own unique value of $\pi$. Taken to its logical conclusion, one could adopt a different value of $\pi$ for each band and each object. This is obviously absurd, but it illustrates the flaws inherent in the reasoning used in W2018.

## 6.3 Linking $p_{IR}$ to $p_v$ creates a false constraint

The assumption in W2018 that $p_v = p_{IR} = p$ reflects a related assumption in some NEOWISE studies. NEOWISE employed 10 different models, at least one of which assumed that $p_v = a\, p_{IR}$, for some slope constant $a$, where $a$ was given different values across the major NEOWISE papers (M2018b). M2018b section 3.3 discussed the rationale for the different values of $a$, which was based on each paper analyzing a different set of asteroids: those in the Main Belt, Jovian Trojans, Hildas, NEOs, and so forth. M2018b discussed this in section 3.3, and M2018b Table 6 listed the assumed values of $a$ for each NEOWISE paper. M2018b computed values of $a$ obtained from linear fits to the asteroids that fit $p_{IR}$ and found them to differ significantly from the values of $a$ that had been chosen in the NEOWISE papers. M2018b concluded that the NEOWISE assumption that $p_v = a\, p_{IR}$ was empirically unsupported.

However, in addition to being empirically a bad assumption or approximation, a less obvious consequence is that it creates a false constraint by linking $p_{IR}$ which depends on the W1, W2 band observations, to $p_v$. This is more problematic for the model than assuming $p_{IR1} = p_{IR2}$ because a spurious term was added to the model for predicting $H$ from $p_v$. The net effect is to gather observational noise from *WISE* band observations and to use that noise to fit $H$.

M2018b showed that many of the $H$ values used in the NEOWISE papers do not match the current values from the MPC. It found that applying 2017 values for $H$ from the MPC to the asteroids in Masiero et al. (2011) changed the resulting value of $p_v$ by 10% or more for 55% of the NEOWISE results. In extreme cases, $p_v$ changed by a factor of up to 43.66. At the time M2018b was written, it seemed that the difference in $H$ values resulted from changes in the MPC, as routinely occurs when new observations are posted. However, it now appears from the discussion in W2018 that the NEOWISE practice of fitting $H$ values may be a major factor as well.

One direct consequence of this issue is that the ~15,000 NEOWISE results that violate Equation (5) are almost certainly in error. Further work is needed to determine the extent and degree of those errors.

## 6.4 Fixed $\eta$ is not an issue

W2018 seizes on a passing note in the appendix of M2018b to make a straw-man argument against treating the beaming parameter $\eta$ as a fixed property of an asteroid. In fact, M2018b notes that $\eta$ can be a function of phase angle. Moreover, Myhrvold (2018a) devotes an entire section to the interpretation of the $\eta$ parameter, which explicitly includes discussion of observation-dependent effects.

To clarify, both Myhrvold (2018a) and M2018b agree that $\eta$ can be a function of phase angle and other factors. However, it is also the case that some studies, such as Harris & Drube (2014, 2016), have argued that $\eta$ nevertheless can often be used to diagnose permanent physical properties of an asteroid, such as whether it is rich in metals.

The issue of $\eta$ is interesting insofar as it relates to the binning of observations into 3- to 10-day epochs of arbitrary duration in the NEOWISE studies. This approach has not been adopted by other thermal-modeling studies. The NEOWISE papers offered no justification for this methodological complication. W2018 now explains that binning was done to deal with phase-angle dependencies in $\eta$.

Neither W2018 nor previous NEOWISE papers have plotted the dependence of $\eta$ and phase angle to demonstrate that the NEOWISE epochs explicitly group by phase angle as claimed. To the contrary, the NEOWISE papers appeared to use epochs based on what the *WISE* observing cadence happened to produce, without regard to phase-angle differences (see M2018b, Table S4 for a summary across NEOWISE papers). The observing geometry of *WISE* and the short duration of the fully cryogenic mission both limit the range of phase angles observed, especially for the main-belt objects that make up the vast majority of NEOWISE asteroids.

The use of arbitrary epochs has serious drawbacks, including the highly deleterious consequence that it forces the NEOWISE team to discard enormous amounts of potentially useful data. That data loss is exacerbated by other data pruning rules (M2018b). Of the raw *WISE* observations, for example, 23% include data from all 4 bands. However, after being cut into epochs and then subjected to various other ad hoc data rules, only about 3% of the NEOWISE results make use of all 4 bands (M2018b). W2018 offers no explanation to rebut the criticisms of these procedures in M2018b. There is a tremendous opportunity for smarter analysis to use more of the precious observational data than NEOWISE has done to date.

Ironically, while W2018 argues that $\eta$ is not fixed, the fact remains that the *majority* of NEOWISE results were obtained by models that not only assume that $\eta$ is fixed for an individual asteroid but also assume it to be fixed across tens of thousands of asteroids. M2018b notes in a lengthy discussion of this issue that it occurs because 5 of the 10 models that NEOWISE uses to estimate diameter do not fit the beaming parameter $\eta$ and instead assign it a fixed value across a set of asteroids. Table 4 of M2018b shows that 55.1% of the NEOWISE results from the PDS use these models. W2018 does not defend the practice and indeed implicitly criticizes it.

The NEOWISE papers typically stated that the fixed value for $\eta$ used for such cases is the mean value found for the same set of asteroids in cases where there is sufficient data to fit $\eta$. However, Table 5 of M2018b found that the value chosen was *not* the mean value for the set as claimed. Moreover, the assumption of fixed $\eta$ was subject to exceptions in some NEOWISE papers, as different values were apparently used for undisclosed reasons.

## 7. The NEOWISE accuracy claims

At several points, W2018 relies heavily on the heuristic that diameter error is half of flux error. This erroneous notion appears as well in NEOWISE papers such as ApJ 736:

> Since diameter is proportional to the square root of the thermal flux (Equation (1)), the minimum systematic diameter error due to uncertainties in the color correction is proportional to one-half the error in flux. These magnitude errors result in a minimum systematic error of ~5%–10% for diameters derived from *WISE* data

As explained in M2018b and here in section 2.1, the heuristic is incorrect. W2018 makes no attempt to rebut the arguments in M2018b that show it to be false.

"Systematic error"—synonymous with systematic bias—denotes a persistent error that is common to a set of results. Systematic error is the mean error one obtains after averaging over a sufficiently large number of results. In addition to systematic error, one also expects random errors that also occur in each individual case to average out to zero across a sufficiently large data set. The total expected error for a given result is the *sum* of the systematic error and the random errors.

The most reasonable interpretation of the statement in ApJ 736 quoted above is that 10% is a very rough estimate of the *lower bound* of the systematic error. Masiero et al. (2011) reference the statement in that sense:

> We note that the flux calibrations presented by Mainzer et al. (2011b) set a lower limit on the accuracy of computed diameters for sources in the *WISE* data of $\sigma D$ = 10%.

Mainzer et al. (2011d) similarly reported that:

> As described in Mainzer et al. (2011b) and Mainzer et al. (2011c), the minimum diameter error that can be achieved using *WISE* observations is ~ 10%, and the minimum albedo error is ~ 20% of the value of the albedo for objects with more than one *WISE* thermal band for which η can be fitted. For objects with large amplitude light curves, poor H or G measurements, or poor signal to noise measurements in the *WISE* bands, the errors will be higher.

If that is the case, then ApJ 736 should be in complete agreement with M2018b, which finds systematic and random errors that are generally higher than 10%, hence *fully consistent* with the ApJ 736 *lower bound*.

Nevertheless, multiple NEOWISE papers, now including W2018, have persisted in exaggerating the accuracy by mischaracterizing the lower bound on systematic error as the total expected error. To cite one prominent example, Masiero et al. (2011) correctly referenced the ApJ 736 results, yet nevertheless claimed in their abstract to have computed diameters of "over 100,000" asteroids "with errors better than 10%."

In fact, the only error analysis offered in Masiero et al. (2011) was a Monte Carlo sensitivity analysis that attempted to determine the error in diameter and other parameters given the *WISE* pipeline estimate uncertainty and uncertainty in other parameters, including *H* and the HG system slope parameter *G*. This would properly called the random error, and exists over and above any systematic error.

M2018b criticized the Masiero et al. (2011) error analysis as inadequate in several ways. Only 25 Monte Carlo trials, an insufficient number, were done for each asteroid. Moreover, as discussed in section 5, the analysis assumed that the *WISE* pipeline $\sigma$ is the standard deviation of a normal distribution for the *WISE* flux errors; it thus grossly underestimated the actual distribution of observational errors of flux. It also failed to include light-curve sampling error (sections 3.5, 4.1). M2018b demonstrated that the net consequence of these flaws is that the per-result Monte Carlo sensitivity analysis greatly underestimated the actual errors, whether systematic or random. W2018 does not rebut that conclusion.

NEOWISE data deposited in the PDS contain model codes that describe the modeling used to obtain each result. Myhrvold (2018a, 2018b) reported that the 10 distinct models used in the NEOWISE data analysis employed 45 different combinations of model and available *WISE* bands. The best combination of model and band (i.e., full thermal modeling for 4-band data, model code DVBI) was used to generate results for fewer than 3,000 asteroids comprising less than 3% of the NEOWISE results.

M2018b also first reported that *more than half* of the NEOWISE results were computed using a single band of data. Neither Masiero et al. (2011), nor any of the NEOWISE studies, presented this overall statistic. Results obtained from a single band were effectively obtained simply by scaling a single hypothetical model asteroid and should not be characterized as products of thermal modeling. W2018 does not rebut or comment on this finding or the conclusion in M2018b that it greatly misrepresents the accuracy of the NEOWISE results to imply that the various model/band combinations employed by NEOWISE all have the same systematic error, as when the quoted passage claimed that "more than 100,000" diameters were estimated to "better than 10%." The actual NEOWISE error analysis supports "worse than 10%," because it is explicitly a *lower bound*. Even that lower bound is claimed for only the best modeling cases, which are fewer than 3% of the overall results (M2018b).

Many other NEOWISE statements have followed this pattern. Grav et al. (2012) asserted, for example, that:

> The *WISE* results have been extensively calibrated against asteroids with known diameter in Mainzer et al. (2011d) and a comparison with the *IRAS* sample is found in Mainzer et al. (2011c). The latter paper shows that the *IRAS*-based diameter values derived by Ryan &Woodward (2010) appear to be systematically larger than diameters from radar, occultation, or spacecraft flybys.

(In this passage Mainzer et al. (2011d) is ApJ 736, and Mainzer et al. (2011c) is ApJ 737.) Comparing one flux value per object for 48 asteroids, which is what was reported in ApJ 736, hardly qualifies as "extensive calibration."

Mainzer et al. (2012) claimed that:

> When two or more thermally-dominated bands are available, η can be estimated, and this was done for more than a hundred thousand asteroids observed by *WISE* during the FC survey phase (Mainzer et al. 2011c; Masiero et al. 2011; Grav et al. 2011, 2012).

(Here Mainzer et al. (2011b) is ApJ 736, and Masiero et al. (2011) has the same reference as in this work.) As M2018b showed, the procedure described was clearly *not* done for more than 100,000 asteroids because many fewer than that number of cases were analyzed using two or more thermally dominated bands.

Exaggerated claims of ~10% accuracy were also made in more recent papers, including Masiero et al. (2013):

> From the NEOWISE survey we now have measurements of diameters for over 130, 000 Main Belt asteroids with relative errors of ~ 10% (see Mas11 and Mainzer et al. 2011b).

and Masiero et al. (2014):

> NEATM provides a rapid method of determining diameter from thermal emission data that is reliable to ~ 10% when the beaming parameter can be fit (Mainzer et al. 2011b).

Note that the condition "when the beaming parameter can be fit" is not a criterion actually studied in ApJ 736 (referenced above as Mainzer et al. (2011b)). One needs to fit the NEATM to determine the beaming parameter $\eta$. The ApJ 736 paper made no claim to do model fitting at all and instead compared fluxes.

In a 2015 review paper, Mainzer et al. (2015) offered more qualifiers to the accuracy claim:

> NEATM-derived diameters generally reproduce measurements from radar, stellar occultations, and *in situ* spacecraft visits to within ±10%, given multiple thermally dominated IR measurements that adequately sample an asteroid's rotational light curve with good signal-to-noise ratio (SNR) and an accurate determination of distance from knowledge of its orbit (Mainzer et al. 2011c). It is worth noting that the accuracy of the diameters of objects used to confirm the performance of radiometric thermal models (such as radar or stellar occultations) is typically ~10%.

But the reference there to Mainzer et al. (2011c) falsely suggests that ApJ 736 presented an analysis of that detailed set of conditions. It did not. Moreover, ApJ 736 stated that the accuracy has a *lower* bound of 10%, not an *upper* bound of 10% as claimed by "within ±10%."

W2018 seeks to redeem the NEOWISE accuracy claims on the basis of two comparison sets presented in its Figure 1. The first is a set of 39 asteroids which had actual modeled diameter estimates (as opposed to copied ROS diameters) published in Masiero et al. 2011. These are compared to ROS diameter estimates published in 2012 and onward.

The results presented by W2018 actually *refute* the $\pm10\%$ accuracy claim—they find that the raw diameter ratios have 16[th] and 84[th] percentiles of 0.93 and 1.15, with a median of 1.04. That implies $1\sigma$ diameter estimates of $-7\%$ to $+15\%$., which is emphatically *not* within $\pm10\%$.

Of course, these figures suffer from the problems recounted above in sections 2 and 3. It is unclear how much they depend on the saturation adjustment to W3, or how applicable they are to typical NEOWISE results. The analysis of W2018 Figure 1 does not account for uncertainty in the NEOWISE or ROS estimates, and they may also average multiple estimates before comparison. The actual numbers are almost certainly less favorable.

The second dataset in W2018 Figure 1 is 123 ROS objects that are found in Masiero et al. 2014. This performs slightly better, with 16[th] and 84[th] percentiles of 0.91 and 1.09, with a median of 0.97. This does come within $\pm10\%$, but is subject to the issues discussed in sections 2 and 3. In addition the Masiero et al. 2014 paper performs only the best level of modeling (see M2018b), which represents only 3% of all NEOWISE results.

Not shown in W2018 is the most important dataset for restoring confidence in the $\pm10\%$: the 117 objects for which NEOWISE papers published copied ROS diameters, along with the 2011 code modeling their diameters. This comparison should have been made in the ApJ 737 paper, but could have been included here.

As shown in M2018b and further developed here, there was never a reason to believe the NEOWISE accuracy claims. They were developed on an extremely small data set (1 flux sample per object per band) with a fundamentally unsound analysis (the scaling of diameter error with flux error). Since then, they have been maintained and exaggerated through mis-referencing and misinterpretation. W2018 does not improve matters by pointedly avoiding the comparison data set that might allow us to settle the matter.

Finally, it should be noted as well that the *accuracy* of most ROS estimates remains undetermined. ROS estimates typically have associated measurement uncertainty estimates for the radar and occultation observations, but those are internal (in-model) measures. Occultation-based measurements are subject to the light-curve effect (section 3.5), as are radar estimates, depending on how many rotational phases of the asteroid are sampled. Direct observations by spacecraft are required to validate those estimates.

# 8. Conclusion

M2018b identified a lengthy set of substantial problems with the NEOWISE assumptions, methods, and results. These flaws have thwarted attempts by other researchers to reproduce the NEOWISE results from the original observations. They also impact the quality and validity of the NEOWISE results, which have been used by hundreds of follow-on studies throughout the planetary science community.

In response, W2018 now brings more clarity to some of the aspects of the NEOWISE study that have long been undisclosed. W2018 confirms many of the most serious problems in NEOWISE results that were identified M2018b. The team has belatedly acknowledged using models that violated Kirchhoff's law of thermal radiation. They admit copying ROS diameters into results tables and presenting those as thermal modeling results. They now acknowledge the existence of a serious bug that corrupted NEOWISE diameter estimates, which was known to them since 2011 but not publicly disclosed until W2018—and that remains unpublished as W2018 was withdrawn after facing criticism during peer review. They have admitted irregularities in their published results, including results for thousands of asteroids being unaccountably added or deleted. They have finally published a saturation correction scheme that was previously undocumented. The absolute magnitudes $H$ that they stated were from the MPC are, in fact, fitted values.

While many of these admitted problems are extremely serious, it is a good that the NEOWISE team now acknowledges them in W2018. For these issues at least, we can now focus on assessing how much damage they have done to the WISE/NEOWISE dataset, and how future work can repair it.

Unfortunately, many of the answers to longstanding questions presented in W2018 raise new questions. The explanation of why the ROS diameters were copied is both inadequate and also in direct contradiction with earlier public statements. The saturation correction scheme has an unexplained origin, and recent work by Wright (2019) confirms that is was developed in the ApJ 736 and ApJ 737 "calibration" papers. Wright (2019) also suggests that the saturation correction was adjusted specifically to make ROS diameters come out correctly. If true, that adjustment negates the applicability of accuracy derived from it to typical cases (which are not saturated), and raises yet another disclosure issue.

The true impact of the software bug is hard to judge because W2018 contradicts itself with different characterizations of the bug: a 6% impact on diameter is either rare and extreme, or typical because $1\sigma$ is $\pm 6\%$.

Something similar occurs with the finding of Hanuš et al. (2015) and Myhrvold (2018b) that the WISE pipeline systematically misestimated observational flux uncertainty. Section 3.1 of W2018 strenuously objects to this result and offers as proof a metric applied to WISE observation of ecliptic polar stars. However, as shown here in section 5, the correct analysis of the ecliptic polar stars upholds the analysis that WISE misestimated uncertainty. Perhaps W2018 agrees: in their conclusion, they contradict all the analysis in their own section 3.1 and state that underestimation of flux uncertainty is a finding of M2018b that has merit—albeit with some qualifications of unknown provenance.

In other areas where W2018 disputes findings of M2018b, they demonstrate a misunderstanding of statistical data analysis. There is no single "correct" or "standard" way to compare thermal modeling diameters with each other, or to ROS diameters. The method advocated by W2018 involves completely ignoring their own NEOWISE estimated diameter uncertainty.

The well known definition of visible band albedo $p_v$ in terms of absolute magnitude *H* and diameter *D* is confused as a reason to fit *H* and treat it as an output of the NEOWISE analysis rather than as an input, despite the fact that the NEATM they claim to use does not predict *H*. This is then justified with a misunderstanding of a basic issue in statistical regression. Extra degrees of freedom *will* be used by the regression, but that does not mean they are valid.

Many of the errors and inadequacies reported by M2018b and now confirmed in W2018 would have been readily caught and corrected years earlier, had NEOWISE fully documented its analytical process, shared its codes, and allowed other researchers to replicate its results in accordance with the normal standards of science. During replication, issues such as the software bug and the misconception that IR observations predict *H* would have come to light sooner. NEOWISE is a case study that illustrates why reproducibility—as well as timely and honest reporting of errors when they are discovered—is so fundamental to the efficient progress of science. Science works best when the community of scientists openly share their work and check each other's results.

Instead, the history of NEOWISE is one of a marked lack of disclosure. Even now, given the opportunity to clarify the record, Wright et al. have regrettably elected not to publish their original pre-bug and post-bug modeled diameters for 117 asteroids for which they copied ROS diameters. Indeed they have chosen not to even *identify* those 117 asteroids. They admit that *H* published in NEOWISE results were actually derived by a fitting process, contrary to the claims in the NEOWISE papers, but they do not disclose how or why their thermal model, claimed to be the NEATM, can "predict" *H*. The linear correction to W3 is now documented, but it appears from Wright (2019) that it was derived as a kind of calibration to make ROS diameters fit better. If that is true, then it represents another serious disclosure omitted from NEOWISE papers and makes the use of the ApJ 736 paper as the basis for NEOWISE accuracy even more problematic than it was previously.

W2018 does not "dispel" misconceptions about NEOWISE, as the title of one of its sections claims. Instead the new revelations and the fundamental misunderstandings behind its failed arguments raise additional concerns about NEOWISE. Rather than rebutting M2018b, W2018 demonstrates why an external and empirical analysis of NEOWISE results was necessary in the first place.

As Myhrvold (2018a) and M2018b emphasized, the observational data set gathered by the *WISE*/NEOWISE missions is potentially one of the greatest treasure troves the asteroid community has yet received. It is a tragedy that its analysis has been plagued by irreproducibility, mismanagement, and secrecy. It should be an urgent priority of the planetary science community to rectify this situation.

# 9. Appendix

*9.1 Sentences referencing ApJ 736*

Two key NEOWISE results papers, Mainzer et al. (2011c) and Masiero et al. (2011), present diameters copied from prior ROS sources in tables of NEOWISE thermally modeled results without disclosing the copying. A list of 105 of these asteroids appears in M2018b, with speculation that there may be more. W2018 discloses that they are actually 117 asteroids with copied diameters across the two papers but does not verify the identity of the 105 asteroids in the M2018b list, nor does it identify the remaining 12 asteroids assuming the 105 are correct.

Masiero (2016) maintained that Masiero et al. (2011) properly referenced this practice, both directly and by referencing the Mainzer et al. (2011c) paper and the ApJ 736 paper "in this respect."

The passages quoted below demonstrate that this assertion is incorrect.

Mainzer et al. (2011c) makes 4 mentions of ApJ 736 (referenced as Mainzer et al. 2011b or M11B in the references for that paper). All 4 are quoted below.

| **Sentences that refer to Mainzer et al. 2011c (here 2011b or M11B) from Mainzer et al. 2011c** |
|---|
| As described in Mainzer et al. (2011b, hereafter M11B) and Cutri et al. (2011), we included observations with magnitudes close to experimentally derived saturation limits, but when sources became brighter than W1 = 6, W2 = 6, W3 = 4, and W4 = 0, we increased the error bars on these points to 0.2 mag and applied a linear correction to W3 (see the *WISE* Explanatory Supplement for details). |
| As described in M11B, we employ the spherical near-Earth asteroid thermal model (NEATM) (Harris 1998). The NEATM model uses the so-called beaming parameter η to account for cases intermediate between zero thermal inertia, the Standard Thermal Model (STM) of Lebofsky & Spencer (1989) and high thermal inertia, the Fast Rotating Model (FRM; Lebofsky et al. 1978; Veeder et al. 1989; Lebofsky & Spencer 1989). |
| The flux from reflected sunlight was computed for each *WISE* band as described in M11B using the IAU phase curve correction (Bowell et al.1989). |
| As described in M11B and Mainzer et al. (2011c), the minimum diameter error that can be achieved using *WISE* observations is ~10%, and the minimum relative albedo error is ~20% for objects with more than one *WISE* thermal band for which η can be fitted. |

To verify the excerpts above, search in the PDF of Mainzer et al. (2011c) for mentions of "M11B."

Masiero et al. (2011) references ApJ 736 (as Mainzer et al. 2011b) in 7 places:

| **Sentences that refer to ApJ 736 (as Mainzer et al. 2011b) results from Masiero et al. 2011** |
|---|
| Prelaunch descriptions of *WISE* were given by Mainzer et al. (2006) and Liu et al. (2008), while postlaunch overviews, including initial calibrations and color corrections, are presented by Wright et al. (2010) and Mainzer et al. (2011b). |
| We obtained our data used for fitting in a method identical to the one described in Mainzer et al. (2011b) and A. K. Mainzer et al. (2011, in preparation), though tuned for MBAs. |
| As discussed in Mainzer et al. (2011b), we find that the pipeline was overly conservative in artifact flagging and cc_flags = p values have similar fluxes to cc_flags = 0 detections while increasing the number of usable observations by ∼20%. |
| Objects with magnitudes brighter than $W3 = 4$ and $W4 = 3$ were assigned errors of 0.2 mag to account for the change in the point-spread function for very bright objects, and a linear correction to the magnitudes of sources with $-2 < W3 < 4$ was applied (Mainzer et al. 2011b; Cutri et al. 2011). |
| We note that the flux calibrations presented by Mainzer et al. (2011b) set a lower limit on the accuracy of computed diameters for sources in the *WISE* data of $\sigma D = 10\%$. |
| Mainzer et al. (2011b) investigated the need for an offset in $H$ to account for systematic errors in $H$ values, but found that no offset was required (cf. Juric et al. 2002, who found a 0.2 mag shift). |
| We perform MC simulations of our visible light measurements as well as of the thermal measurements to quantify the error on albedo; however, in all cases the minimum error on albedo will be 20% (Mainzer et al. 2011b) for objects with optical data and one good thermal band. |

To verify the excerpts above, search in the PDF of Masiero et al. (2011) for mentions of "2011b."

That paper refers to their diameter results in the following excerpts (a pruned subset of the sentences that contain "diameter"):

| **Sentences that refer to diameter results from Masiero et al. 2011** |
|---|
| Using a NEATM thermal model fitting routine, we compute diameters for over 100,000 Main Belt asteroids from their IR thermal flux, with errors better than 10% |
| A large survey conducted in mid-infrared wavelengths will allow Main Belt albedos and diameters to be produced with good accuracy; this in turn will allow us to study the compositional gradient of the solar system and may ultimately allow us to set constraints on any major planetary migration that may have occurred. |
| For objects that have known orbits, measurement of the infrared flux emitted from the surface can be used to constrain the diameter of the body (see Section 3 for a discussion of the method used here). |
| Thermal infrared measurements of a large sample of asteroids represent the best way to determine robust diameters rapidly for many thousands of objects. In this paper, we present preliminary results from the Wide-field Infrared Survey Explorer (WISE) space telescope, the next-generation all-sky infrared survey, focusing here on the cryogenic observations of MBAs. |
| 3. DIAMETER AND ALBEDO DETERMINATION THROUGH THERMAL MODELING (section title) |
| We note that the flux calibrations presented by Mainzer et al. (2011b) set a lower limit on the accuracy of computed diameters for sources in the *WISE* data of σD = 10%. |
| In total 129,750 MBAs, selected from the cryogenic phase of the survey, had a sufficient number and quality of detections to allow us to perform thermal modeling and determine their effective diameter. Of these, 17,482 objects had orbital arcs shorter than 30 days; as such their orbits have a larger uncertainty than the rest of the population, which corresponds to uncertainty in their geocentric and heliocentric distances, which will naturally increase the error on their calculated diameters. Additionally, 112,265 objects also had available optical data allowing us to calculate albedo as well as diameter. |

> We provide the full table of our best fits for MBAs from the Pass 1 processed cryogenic survey data in the online version of the journal, or online at: http://wise2.ipac.caltech.edu/staff/bauer/NEOWISE_pass1/. A sample of the table is shown in Table1.This table contains: the MPC-packed format name of the object; the H and G values used; the diameter, albedo, beaming parameter, and infrared albedo as well as associated error bars; and the number of observations in each *WISE* band that were used for fitting.

> **Table 1** Example of Electronic Table of the Thermal Model Fits (title of example table of NEOWISE results)

> Using fluxes from the *WISE* data, and a faceted NEATM model, we are able to determine diameters for our observed objects.

> With infrared fluxes of sufficient quality to determine diameters and albedos for 129,750 MBAs, we show the power and great potential contained in this data set.

None of the statements above is consistent with choosing ROS diameters for more than 100 asteroids in the data set. On the contrary, each statement sets the clear expectation that the diameter results were obtained by thermal modeling.

## 9.2 Tables of distribution fit parameters

| Name | W1 | | W2 | | W3 | | W4 | |
|---|---|---|---|---|---|---|---|---|
| | μ | σ | μ | σ | μ | σ | μ | σ |
| NGC6552 | | | 0.017 | 1.427 | −0.038 | 1.237 | −0.163 | 1.119 |
| KF03T1 | −1.364 | 5.973 | −1.318 | 2.491 | −0.181 | 1.495 | −0.384 | 3.334 |
| KF03T2 | −0.027 | 1.067 | −0.089 | 1.542 | −0.134 | 1.429 | −0.408 | 3.025 |
| KF06T1 | −0.124 | 2.279 | −0.114 | 1.373 | −0.364 | 1.502 | −0.283 | 2.749 |
| KF06T2 | −0.164 | 1.231 | −0.014 | 1.196 | −0.316 | 1.475 | −0.299 | 2.613 |
| KF06T3 | −1.260 | 1.633 | −0.219 | 1.373 | −0.304 | 1.638 | −0.444 | 3.380 |
| KF03T4 | −0.289 | 1.724 | −0.018 | 1.432 | −0.209 | 1.465 | −0.291 | 2.510 |
| KF05T1 | −0.049 | 1.054 | −0.017 | 1.885 | −0.109 | 1.553 | | |
| KF02T1 | −0.052 | 1.309 | −0.033 | 2.164 | −0.068 | 1.429 | −0.232 | 1.987 |
| Bp66_1073 | −0.008 | 1.573 | −0.005 | 1.518 | −0.066 | 1.256 | −0.292 | 1.467 |
| KF02T3 | −1.002 | 1.834 | −0.374 | 1.771 | −0.209 | 1.441 | −0.327 | 2.781 |
| HD270422 | 0.033 | 1.794 | −0.006 | 1.588 | −0.028 | 1.243 | −0.262 | 1.237 |
| HD270467 | −0.040 | 1.210 | −0.010 | 1.591 | −0.045 | 1.248 | −0.158 | 1.346 |
| 5581475 | −0.078 | 1.313 | −0.070 | 1.382 | −0.139 | 1.075 | −0.489 | 3.543 |
| WHO_G642 | −0.537 | 1.624 | −0.060 | 1.651 | −0.083 | 1.215 | −0.277 | 2.223 |
| HD41466 | −0.018 | 1.165 | −0.010 | 1.508 | −0.025 | 1.323 | −0.180 | 1.362 |
| WHO_S527 | −0.015 | 1.566 | −0.012 | 1.737 | −0.018 | 1.119 | −0.241 | 1.207 |
| HD270485 | −0.169 | 2.732 | −0.212 | 2.170 | −0.087 | 1.255 | −0.258 | 2.203 |
| HD271776 | −0.147 | 1.310 | −0.102 | 1.544 | −0.108 | 1.211 | −0.367 | 3.210 |

**Table A1. Parameters of best-fit normal distributions obtained by maximum likelihood estimation.**
The best-fit normal distribution to the Z-statistic of Equation (1) for repeated observations of the sources of Table 1 is shown by source, band, and parameter. Cases having fewer than 100 eligible observations after data filtering, as described in the text, were not fit; those entries are blank. The standard deviation is close to unity for only one source and band (the W1 band for star KF05T1). All other cases show significant underestimation of uncertainty by the *WISE* pipeline $\sigma$.

| Name | W1 | | | W2 | | | W3 | | | W4 | | |
|---|---|---|---|---|---|---|---|---|---|---|---|---|
| | μ | σ | ν | μ | σ | ν | μ | σ | ν | μ | σ | ν |
| NGC6552 | | | | 0.031 | 1.304 | 12.516 | −0.008 | 1.029 | 6.326 | −0.088 | 0.779 | 3.764 |
| KF03T1 | −1.795 | 0.747 | 2.368 | −1.486 | 1.066 | 3.271 | −0.066 | 1.201 | 6.060 | 0.552 | 0.285 | 1.000 |
| KF03T2 | −0.046 | 0.806 | 4.822 | −0.093 | 1.194 | 6.156 | −0.110 | 1.286 | 11.005 | 0.414 | 0.383 | 1.241 |
| KF06T1 | −0.182 | 0.923 | 4.652 | −0.125 | 1.036 | 6.791 | −0.108 | 1.099 | 4.320 | 0.812 | 0.259 | 0.814 |
| KF06T2 | −0.216 | 0.816 | 5.020 | −0.030 | 1.092 | 12.177 | 0.103 | 0.763 | 2.331 | 0.777 | 0.217 | 0.749 |
| KF06T3 | −1.397 | 0.825 | 3.252 | −0.230 | 1.034 | 5.358 | −0.111 | 1.295 | 5.493 | 0.991 | 0.218 | 0.712 |
| KF03T4 | −0.369 | 0.792 | 4.273 | −0.011 | 1.095 | 6.051 | −0.130 | 1.274 | 8.711 | 0.643 | 0.285 | 0.922 |
| KF05T1 | −0.065 | 0.818 | 5.087 | 0.002 | 1.129 | 5.507 | −0.068 | 1.345 | 8.236 | | | |
| KF02T1 | −0.100 | 0.799 | 4.133 | 0.053 | 1.005 | 3.635 | −0.029 | 1.247 | 8.927 | 0.289 | 0.431 | 1.599 |
| Bp66_1073 | −0.072 | 1.315 | 6.584 | 0.007 | 1.280 | 7.044 | −0.060 | 1.113 | 9.648 | 0.011 | 0.813 | 3.120 |
| KF02T3 | −1.105 | 0.949 | 3.616 | −0.364 | 1.000 | 5.050 | −0.103 | 1.213 | 7.283 | 0.629 | 0.202 | 0.832 |
| HD270422 | −0.049 | 1.474 | 6.056 | 0.031 | 1.289 | 6.249 | −0.024 | 1.149 | 14.460 | −0.087 | 0.887 | 4.533 |
| HD270467 | −0.065 | 0.801 | 4.074 | 0.016 | 1.104 | 5.677 | −0.040 | 1.072 | 8.312 | 0.228 | 0.309 | 1.503 |
| 5581475 | −0.123 | 0.900 | 5.259 | −0.086 | 1.094 | 6.794 | 0.154 | 0.346 | 1.827 | 0.802 | 0.240 | 0.806 |
| WHO_G642 | −0.607 | 0.710 | 3.455 | 0.000 | 1.154 | 4.954 | −0.054 | 1.073 | 9.785 | 0.400 | 0.273 | 1.115 |
| HD41466 | −0.037 | 0.882 | 5.070 | 0.036 | 1.075 | 5.108 | −0.037 | 1.147 | 8.536 | 0.158 | 0.412 | 1.903 |
| WHO_S527 | −0.048 | 1.186 | 4.464 | 0.001 | 1.300 | 5.506 | −0.021 | 0.966 | 8.047 | −0.028 | 0.756 | 3.663 |
| HD270485 | −0.203 | 0.896 | 4.113 | −0.190 | 1.135 | 4.145 | −0.075 | 1.097 | 8.678 | 0.343 | 0.248 | 1.132 |
| HD271776 | −0.108 | 0.829 | 3.843 | −0.077 | 1.138 | 5.458 | −0.065 | 1.044 | 8.697 | 0.479 | 0.237 | 1.056 |

**Table A2. Parameters of best-fit Student's t-distribution obtained by maximum likelihood estimation.** The best-fit Student's t-distribution to the Z-statistic of Equation (1) for repeated observations of the sources of Table 1 is shown by source, band, and parameter. Cases having fewer than 100 eligible observations after data filtering, as described in the text, were not fit; those entries are blank.

# 10. References

Andrec, M., Kholodenko, B.N., Levy, R.M., Sontag, E., 2005. Inference of signaling and gene regulatory networks by steady-state perturbation experiments: structure and accuracy. J. Theor. Biol. 232, 427–441.

Arbutina, D.S., Kovačević, U.D., 2019. Probabilistic Analysis of Voltage Divider Ratios. FME Trans. 47, 625.

Bevington, P.R., Robinson, D.K., 2003. Data Reduction and Error Analysis for the Physical Sciences, 3rd ed. McGraw-Hill, New York, N.Y.

Bonamente, M., 2013. Statistics and Analysis of Scientific Data, 2nd ed. Springer, New York, N.Y.

Buchheim, R.K., 2010. Methods and lessons learned determining the hg parameters of asteroid phase curves, in: Society for Astronomical Sciences Annual Symposium. pp. 101–115.

Clemon, L.M., Zohdi, T.I., 2018. On the tolerable limits of granulated recycled material additives to maintain structural integrity. Constr. Build. Mater. 167, 846–852.

Cutri, R.M., Wright, E.L., Conrow, T., Bauer, J.M., Benford, D.J., Brandenburg, H., Dailey, J., Eisenhardt, P.R.M., Evans, T., Fajardo-Acosta, S.B., Fowler, J., Gelino, C., Grillmair, C., Harbut, M., Hoffman, D., Jarrett, T.H., Kirkpatrick, J.D., Liu, W., Mainzer, A., Marsh, K., Masci, F., McCallon, H., Padgett, D., Ressler, M.E., Royer, D., Skrutskie, M.F., Stanford, S.A., Wyatt, P.L., Tholen, D., Tsai, C.-W., Wachter, S., Wheelock, S.L., Yan, L., Alles, R., Beck, R., Grav, T., Masiero, J., McCollum, B., McGehee, P., Wittman, M., 2011. Explanatory Supplement to the WISE Preliminary Data Release Products [WWW Document]. Explan. Suppl. to WISE Prelim. Data Release Prod. URL http://wise2.ipac.caltech.edu/docs/release/prelim/expsup/

Delbó, M., Dell'Oro, A., Harris, A.W., Mottola, S., Mueller, M., 2007. Thermal inertia of near-Earth asteroids and implications for the magnitude of the Yarkovsky effect. Icarus 190, 236–249. https://doi.org/10.1016/j.icarus.2007.03.007

Delbo, M., Mueller, M., Emery, J.P., Rozitis, B., Capria, M.T., 2015. Asteroid Thermophysical Modeling, in: Asteroids IV. University of Arizona Press, Tucson, AZ, pp. 107–128. https://doi.org/10.2458/azu_uapress_9780816532131-ch006

Dhanoa, M.S., Sanderson, R., Shanmugalingam, S., López, S., Murray, J.M.D., France, J., 2018. The distribution of the ratio of two correlated measured variables may not always be normal: Case studies related to meat quality and animal nutrition. e-planet 16.

Dieck, R.H., 2017. Measurement uncertainty: methods and applications, 5th ed. ISA.

Ďurech, J., Sidorin, V., Kaasalainen, M., 2010. DAMIT: a database of asteroid models. Astron. Astrophys. 513, A46. https://doi.org/10.1051/0004-6361/200912693

Ezzamel, M., Mar-Molinero, C., Beech, A., 1987. On the Distributional Properties of Financial Ratios. J. Bus. Financ. Account. 14, 463–481. https://doi.org/10.1111/j.1468-5957.1987.tb00107.x

Feigelson, E.D., Babu, G.J., 2013. Statistical Methods for Astronomy, in: Planets, Stars and Stellar Systems. Springer Netherlands, pp. 445–480.

Feigelson, E.D., Babu, G.J., 2012. Modern Statistical Methods for Astronomy: With R

Applications, 1st ed. Cambridge University Press.

Frühwirth-Schnatter, S., 2006. Finite Mixture and Markov Switching Models (Springer Series in Statistics). Springer.

González-Marcos, A., Sarro, L.M., Ordieres-Meré, J., Bello-García, A., 2016. Evaluation of data compression techniques for the inference of stellar atmospheric parameters from high-resolution spectra. Mon. Not. R. Astron. Soc. 465, 4556–4571. https://doi.org/10.1093/mnras/stw3031

Grav, T., Mainzer, A., Bauer, J., Masiero, J., Spahr, T., McMillan, R.S., Walker, R., Cutri, R., Wright, E., Eisenhardt, P.R.M., Blauvelt, E., DeBaun, E., Elsbury, D., Gautier, T., Gomillion, S., Hand, E., Wilkins, A., 2011a. WISE/NEOWISE Observations of the Jovian Trojans: Preliminary Results. Astrophys. J. 742, 40. https://doi.org/10.1088/0004-637X/742/1/40

Grav, T., Mainzer, A., Bauer, J.M., Masiero, J., Spahr, T., McMillan, R.S., Walker, R., Cutri, R., Wright, E.L., Eisenhardt, P.R.M., Blauvelt, E., DeBaun, E., Elsbury, D., Gautier, T., Gomillion, S., Hand, E., Wilkins, A., 2011b. WISE/Neowise Observations of the Hilda Population: Preliminary Results. Astrophys. J. 744, 197. https://doi.org/10.1088/0004-637X/744/2/197

Grav, T., Mainzer, A., Bauer, J.M., Masiero, J.R., Nugent, C.R., 2012. WISE/NEOWISE Observations of the Jovian Trojan Population: Taxonomy. Astrophys. J. 759, 49. https://doi.org/10.1088/0004-637X/759/1/49

Hanuš, J., Delbó, M., Ďurech, J., Alí-Lagoa, V., 2015. Thermophysical modeling of asteroids from WISE thermal infrared data – Significance of the shape model and the pole orientation uncertainties. Icarus 256, 101–116. https://doi.org/10.1016/j.icarus.2015.04.014

Harris, A.W., 2005. The surface properties of small asteroids from thermal-infrared observations. Proc. Int. Astron. Union 1, 449–463. https://doi.org/10.1017/S1743921305006915

Harris, A.W., 1998. A Thermal Model for Near-Earth Asteroids. Icarus 131, 291–301. https://doi.org/10.1006/icar.1997.5865

Harris, A.W., Drube, L., 2016. Thermal tomography of asteroid surface structure. Astrophys. J. 832, 127. https://doi.org/10.3847/0004-637X/832/2/127

Harris, A.W., Drube, L., 2014. How To Find Metal-Rich Asteroids. Astrophys. J. 785, L4. https://doi.org/10.1088/2041-8205/785/1/L4

Harris, A.W., Harris, A.W., 1997. On the revision of radiometric albedos and diameters of asteroids. Icarus 126, 450–454. https://doi.org/10.1006/icar.1996.5664

Harris, A.W., Lagerros, J.S.V., 2002. Asteroids in the Thermal Infrared, in: Asteroids III. University of Arizona Press, Tucson, pp. 205–218.

Harrison, F., 2011. Getting started with meta-analysis. Methods Ecol. Evol. 2, 1–10. https://doi.org/10.1111/j.2041-210X.2010.00056.x

Hedges, L. V., Gurevitch, J. and Curtis, P.S., 1999. The Meta-Analysis of Response Ratios in Experimental Ecology. Ecology 80, 1150–1156. https://doi.org/10.1890/0012-9658(1999)080[1150:TMAORR]2.0.CO;2

Hinkley, D. V, 1969. On the ratio of two correlated normal random variables. Biometrika 56, 635–639.

Israel, H., Kitching, T. D., Massey, R., Cropper, M., 2017. Problems using ratios of galaxy shape moments in requirements for weak lensing surveys. A&A 598, A46. https://doi.org/10.1051/0004-6361/201628394

Ivezić, Ž., Connolly, A., VanderPlas, J., Gray, A., 2014. Statistics, Data Mining, and Machine Learning in Astronomy: A Practical Python Guide for the Analysis of Survey Data (Princeton Series in Modern Observational Astronomy). Princeton University Press, Princeton, N. J.

Jarrett, T.H., Cohen, M., Masci, F., Wright, E.L., Stern, D., Benford, D.J., Blain, A.W., Carey, S.J., Cutri, R.M., Eisenhardt, P.R.M., Lonsdale, C., Mainzer, A., Marsh, K., Padgett, D., Petty, S., Ressler, M., Skrutskie, M., Stanford, S., Surace, J., Tsai, C.W., Wheelock, S., Yan, D.L., 2011. The Spitzer - Wise Survey of the Ecliptic Poles. Astrophys. J. 735, 112. https://doi.org/10.1088/0004-637X/735/2/112

Kim, S., Lee, H.M., Nakagawa, T., Hasegawa, S., 2003. Thermal Models and Far Infrared Emission of Asteroids. J. Korean Astron. Soc. 36, 21–31. https://doi.org/10.5303/JKAS.2003.36.1.021

Kordas, G., Petrakos, G., 2017. The Distribution of the Ratio of Normal Random Variables and the Ellipticity of the Earth. Metod. Zv. 14, 19–35.

Koren, S.C., Wright, E.L., Mainzer, A., 2015. Characterizing asteroids multiply-observed at infrared wavelengths. Icarus 258, 82–91. https://doi.org/10.1016/j.icarus.2015.06.014

Korhonen, P.J., Narula, S.C., 1989. The probability distribution of the ratio of the absolute values of two normal variables. J. Stat. Comput. Simul. 33, 173–182.

Lian, J., Zhu, Q., Kong, X., He, J., 2014. Characterizing AGB stars in Wide-field Infrared Survey Explorer (WISE) bands. Astron. Astrophys. 564, A84. https://doi.org/10.1051/0004-6361/201322818

Ly, Sel, Pho, K.-H., Ly, Sal, Wong, W.-K., 2019. Determining Distribution for the Quotients of Dependent and Independent Random Variables by Using Copulas. J. Risk Financ. Manag. 12. https://doi.org/10.3390/jrfm12010042

Mainzer, A., Bauer, J.M., Cutri, R.M., Grav, T., Kramer, E.A., Masiero, J., Nugent, C.R., Sonnett, S.M., Stevenson, R.A., Wright, E.L., 2016. NEOWISE Diameters and Albedos V1.0. EAR-A-COMPIL-5-NEOWISEDIAM-V1.0. NASA Planet. Data Syst.

Mainzer, A., Bauer, J.M., Grav, T., Masiero, J., Cutri, R.M., Wright, E., Nugent, C.R., Stevenson, R., Clyne, E., Cukrov, G., Masci, F., 2014. The Population of Tiny Near-Earth Objects Observed by NEOWISE. Astrophys. J. 784, 110. https://doi.org/10.1088/0004-637X/784/2/110

Mainzer, A., Grav, T., Bauer, J.M., Masiero, J., McMillan, R.S., Cutri, R.M., Walker, R., Wright, E.L., Eisenhardt, P.R.M., Tholen, D.J., Spahr, T., Jedicke, R., Denneau, L., DeBaun, E., Elsbury, D., Gautier, T., Gomillion, S., Hand, E., Mo, W., Watkins, J., Wilkins, A., Bryngelson, G.L., Del Pino Molina, A., Desai, S., Camus, M.G., Hidalgo, S.L., Konstantopoulos, I., Larsen, J.A., Maleszewski, C., Malkan, M.A., Mauduit, J.-C., Mullan, B.L., Olszewski, E.W., Pforr, J., Saro, A., Scotti, J. V., Wasserman, L.H., 2011a. Neowise Observations of Near-Earth Objects: Preliminary Results. Astrophys. J. 743, 156. https://doi.org/10.1088/0004-637X/743/2/156

Mainzer, A., Grav, T., Masiero, J., Bauer, J.M., Cutri, R.M., Mcmillan, R.S., Nugent, C.R., Tholen, D., Walker, R., Wright, E.L., 2012. Physical Parameters of Asteroids Estimated from the WISE 3 Band Data and NEOWISE Post-Cryogenic Survey. Astrophys. J. Lett. 760, L12. https://doi.org/10.1088/2041-8205/760/1/L12

Mainzer, A., Grav, T., Masiero, J., Bauer, J.M., Wright, E.L., Cutri, R.M., McMillan, R.S., Cohen, M., Ressler, M., Eisenhardt, P.R.M., 2011b. Thermal Model Calibration for Minor Planets Observed With Wide-Field Infrared Survey Explorer/Neowise. Astrophys. J. 736, 100. https://doi.org/10.1088/0004-637X/736/2/100

Mainzer, A., Grav, T., Masiero, J., Bauer, J.M., Wright, E.L., Cutri, R.M., Walker, R., McMillan, R.S., 2011c. Thermal Model Calibration for Minor Planets Observed with WISE/NEOWISE: Comparison with Infrared Astronomical Satellite. Astrophys. J. Lett. 737, L9. https://doi.org/10.1088/2041-8205/737/1/L9

Mainzer, A., Grav, T., Masiero, J., Hand, E., Bauer, J.M., Tholen, D., McMillan, R.S., Spahr, T., Cutri, R.M., Wright, E.L., Watkins, J., Mo, W., Maleszewski, C., 2011d. NEOWISE studies of spectrophotometrically classified asteroids: Preliminary results. Astrophys. J. 741, 90. https://doi.org/10.1088/0004-637X/741/2/90

Mainzer, A., Usui, F., Trilling, D.E., 2015. Space-Based Thermal Infrared Studies of Asteroids, in: Asteroids IV. University of Arizona Press, Tucson, AZ, p. 89.

Mandel, I., Farr, W.M., Gair, J.R., 2019. Extracting distribution parameters from multiple uncertain observations with selection biases. Mon. Not. R. Astron. Soc. 486, 1086–1093. https://doi.org/10.1093/mnras/stz896

Mandel, J., 1964. The Statistical Analysis of Experimental Data. Dover.

Marsaglia, G., 1965. Ratios of normal variables and ratios of sums of uniform variables. J. Am. Stat. Assoc. 60, 193–204.

Masiero, J., 2016. Maisiero Post to Yahoo minor planets forum. https://groups.yahoo.com/neo/groups/mpml/conversations/messages/32047

Masiero, J., Grav, T., Mainzer, A., Nugent, C.R., Bauer, J.M., Stevenson, R., Sonnett, 2014. Main Belt Asteroids with WISE/NEOWISE: Near-Infrared Albedos. Astrophys. J. 791, 121. https://doi.org/10.1088/0004-637X/791/2/121

Masiero, J., Jedicke, R., Durech, J., Gwyn, S., Denneau, L., Larsen, J., 2009. The Thousand Asteroid Light Curve Survey. Icarus 204, 145–171. https://doi.org/10.1016/j.icarus.2009.06.012

Masiero, J., Mainzer, A., Bauer, J.M., Grav, T., Nugent, C.R., Stevenson, R., 2013. Asteroid family identification using the Hierarchical Clustering Method and WISE/NEOWISE physical properties. Astrophys. J. 770, 7. https://doi.org/10.1088/0004-637X/770/1/7

Masiero, J., Mainzer, A., Grav, T., Bauer, J.M., Cutri, R.M., Dailey, J., Eisenhardt, P.R.M., McMillan, R.S., Spahr, T.B., Skrutskie, M.F., Tholen, D., Walker, R.G., Wright, E.L., DeBaun, E., Elsbury, D., Gautier IV, T., Gomillion, S., Wilkins, A., 2011. Main Belt Asteroids with WISE/NEOWISE. I. Preliminary Albedos and Diameters. Astrophys. J. 741, 68. https://doi.org/10.1088/0004-637X/741/2/68

Masiero, J., Mainzer, A., Grav, T., Bauer, J.M., Cutri, R.M., Nugent, C., Cabrera, M.S., 2012. Preliminary analysis of WISE/NEOWISE 3-band cryogenic and post-cryogenic observations of main belt asteroids. Astrophys. J. 759, L8. https://doi.org/10.1088/2041-

8205/759/1/L8

Matson, D.L., 1986. The IRAS asteroid and comet survey, in: Bulletin of the American Astronomical Society. p. 790.

McLachlan, G., Peel, D., 2000. Finite Mixture Models. Wiley-Interscience.

Mekić, E., Stefanović, M., Spalević, P., Sekulović, N., Stanković, A., 2012. Statistical Analysis of Ratio of Random Variables and Its Application in Performance Analysis of Multihop Wireless Transmissions. Math. Probl. Eng. https://doi.org/10.1155/2012/841092

Melchior, P., Viola, M., 2012. Means of confusion: how pixel noise affects shear estimates for weak gravitational lensing. Mon. Not. R. Astron. Soc. 424, 2757–2769. https://doi.org/10.1111/j.1365-2966.2012.21381.x

Myhrvold, N., 2018a. An empirical examination of WISE/NEOWISE asteroid analysis and results. Icarus 314, 64–97. https://doi.org/10.1016/j.icarus.2018.05.004

Myhrvold, N., 2018b. Asteroid thermal modeling in the presence of reflected sunlight. Icarus 303, 91–113. https://doi.org/10.1016/j.icarus.2017.12.024

Nugent, C.R., Mainzer, A., Masiero, J., Wright, E.L., Bauer, J., Grav, T., Kramer, E.A., Sonnett, S., 2016. Observed asteroid surface area in the thermal infrared. Astron. J. 153, 90. https://doi.org/10.3847/1538-3881/153/2/90

Pham-Gia, T., Turkkan, N., Marchand, E., 2006. Density of the ratio of two normal random variables and applications. Commun. Stat. Methods 35, 1569–1591.

Rozitis, B., Green, S.F., 2011. Directional characteristics of thermal-infrared beaming from atmosphereless planetary surfaces—a new thermophysical model. Mon. Not. R. Astron. Soc. 415, 2042–2062. https://doi.org/10.1111/j.1365-2966.2011.18718.x

Ryan, E.L., Woodward, C.E., 2010. Rectified asteroid albedos and diameters from IRAS and MSX Photometry Catalogs. Astron. J. 140, 933. https://doi.org/10.1088/0004-6256/140/4/933

Shevchenko, V.G., Tedesco, E.F., 2006. Asteroid albedos deduced from stellar occultations. Icarus 184, 211–220. https://doi.org/10.1016/j.icarus.2006.04.006

Stuart, A., Ord, J.K., 2010. Kendall's advanced theory of statistics, 6th ed. Wiley.

Swamy, T.J., Avasarala, S., Sandhya, T., Ramamurthy, G., 2012. Spectrum sensing: Approximations for Eigenvalue ratio based detection, in: 2012 International Conference on Computer Communication and Informatics. pp. 1–5. https://doi.org/10.1109/ICCCI.2012.6158914

Taylor, J., 1997. Introduction to error analysis, the study of uncertainties in physical measurements, 2nd ed.

Tedesco, E.F., Noah, P. V., Noah, M., Price, S.D., 2002. The Supplemental IRAS Minor Planet Survey. Astron. J. 123, 1056–1085. https://doi.org/10.1086/338320

Thomaseth, C., Radde, N., 2016. Normalization of western blot data affects the statistics of estimators. IFAC-PapersOnLine 49, 56–62.

Trilling, D.E., Mueller, M., Hora, J.L., Fazio, G., Spahr, T., Stansberry, J.A., Smith, H.A., Chesley, S.R., Mainzer, A., 2008. Diameters and albedos of three sub-kilometer Near Earth

Objects derived from Spitzer observations. Astrophys. J. Lett. 683. https://doi.org/10.1086/591668

Veres, P., Jedicke, R., Denneau, L., 2012. Improved Asteroid Astrometry and Photometry with Trail Fitting. Publ. … 124, 1197–1207. https://doi.org/10.1086/668616

Vereš, P., Jedicke, R., Fitzsimmons, A., Denneau, L., Bolin, B., Wainscoat, R., Tonry, J., 2014. Absolute magnitudes and slope parameters of Pan-STARRS PS1 asteroids—preliminary results, in: Asteroids, Comets, Meteors. Helsinki, p. 1.

Wall, J.V., Jenkins, C.R., 2012. Practical Statistics for Astronomers (Cambridge Observing Handbooks for Research Astronomers), 2nd ed. Cambridge University Press.

Wolters, S.D., Green, S.F., 2009. Investigation of systematic bias in radiometric diameter determination of near-Earth asteroids: The night emission simulated thermal model (NESTM). Mon. Not. R. Astron. Soc. 400, 204–218. https://doi.org/10.1111/j.1365-2966.2009.14996.x

Wright, E., Mainzer, A., Masiero, J., Grav, T., Cutri, R., Bauer, J., 2018. Response to "An empirical examination of WISE/NEOWISE asteroid analysis and results" 1–30. arXiv:1811.01454v1 [astro-ph.EP]

Wright, E.L., 2019. Testing the WISE Saturation Correction with Asteroids, in: American Astronomical Society Meeting Abstracts# 233.

Wright, E.L., Eisenhardt, P.R.M., Mainzer, A., Ressler, M.E., Cutri, R.M., Jarrett, T., Kirkpatrick, J.D., Padgett, D., McMillan, R.S., Skrutskie, M., Stanford, S.A., Cohen, M., Walker, R.G., Mather, J.C., Leisawitz, D., Gautier, T.N., McLean, I., Benford, D.J., Lonsdale, C.J., Blain, A., Mendez, B., Irace, W.R., Duval, V., Liu, F., Royer, D., Heinrichsen, I., Howard, J., Shannon, M., Kendall, M., Walsh, A.L., Larsen, M., Cardon, J.G., Schick, S., Schwalm, M., Abid, M., Fabinsky, B., Naes, L., Tsai, C.-W., 2010. The Wide-field Infrared Survey Explorer (WISE): Mission Description and Initial On-orbit Performance. Astron. J. 140, 1868. https://doi.org/10.1088/0004-6256/140/6/1868